\documentclass[a4paper, onecolumn]{article}

\usepackage[utf8]{inputenc}\DeclareUnicodeCharacter{2212}{-}
\usepackage{biblatex}
\usepackage[affil-it]{authblk}
\usepackage[a4paper, textwidth=16cm, textheight=25cm]{geometry}
\usepackage{amsmath}  
\usepackage{amsfonts} 
\usepackage{nicefrac} 
\usepackage[ruled,vlined]{algorithm2e} 
\usepackage{pgf} 
\usepackage{graphicx,psfrag,color}
\usepackage{float} 
\usepackage[printwatermark]{xwatermark}
\newwatermark*[allpages,angle=90,scale=1.5,xpos=85,ypos=0]{pre-print submitted to Elsevier}
\title{\bf Reconstructing random heterogeneous media through differentiable optimization}
\author[1]{Paul Seibert}
\author[1]{Marreddy Ambati}
\author[1]{Alexander Ra{\ss}loff}
\author[1,2,3]{Markus K\"astner\thanks{{Corresponding Author. \it{E-mail address}:} markus.kaestner@tu-dresden.de (Markus K\"astner)}}

\affil[1]{Institute of Solid Mechanics, TU Dresden,  01062 Dresden, Germany}
\affil[2]{Dresden Center for Computational Materials Science, TU Dresden, 01062 Dresden, Germany}
\affil[3] {Dresden Center for Fatigue and Reliability, 01062 Dresden, Germany}

\date{}
\begin{document}
\maketitle

\hrule

\begin{abstract}
  Microstructure reconstruction is a key enabler of process-structure-property linkages, a central topic in materials engineering. Revisiting classical optimization-based reconstruction techniques, they are recognized as a powerful framework to reconstruct random heterogeneous media, especially due to their generality and controllability. The stochasticity of the available approaches is, however, identified as a performance bottleneck. 
  In this work, reconstruction is approached as a \emph{differentiable} optimization problem, where the error of a generic prescribed descriptor is minimized under consideration of its derivative. As an exemplary descriptor, a suitable differentiable version of spatial correlations is formulated, along with a multigrid scheme to ensure scalability. 
  The applicability of differentiable optimization realized through this descriptor is demonstrated using a wide variety of heterogeneous media, achieving exact statistical equivalence with errors as low as~$0\%$ in a short time. We conclude that, while still in an early stage of development, this approach has the potential to significantly alleviate the computational effort currently associated with reconstructing general random heterogeneous media. 
\end{abstract}
\hrule

\section{Introduction}
\label{sec:intro}
Microstructure characterization and reconstruction (MCR) is an essential component in inverse material design, a concept to accelerate materials engineering~\cite{olson1997, fullwood_microstructure_2010, mcdowell2016, kalidindi2019, bock2019c}. The key for inverse material design is to establish reliable process-structure-property (PSP) linkages. For this purpose, the wide gap between the manufacturing processes, which can often be controlled, and the material properties, which are the primal interest in practical engineering applications, is bridged by the microstructure. Given a number of microstructures, it is desirable to perform comparative analyses on them, like the detection of common cases and rare outliers, or even averaging and interpolating. Due to the general morphological randomness of heterogeneous materials, such comparative analyses require a translation-invariant microstructure description, often called microstructure descriptor or characterization. Such descriptors range from simple physical descriptors like volume fractions~\cite{xu_descriptor-based_2014} to more general approaches~\cite{torquato_statistical_2002} like the lineal-path function~\cite{lu1992}, $n$-point cluster correlation function~\cite{torquato2002} and $n$-point spatial correlation function~\cite{torquato2002, jiao_modeling_2007, kalidindi2011}. Especially the two-point correlation function, a member of the latter group, has proven valuable for various heterogeneous materials~\cite{bostanabad_computational_2018}. The potential of microstructure descriptors, however, does not fully unfold without the reconstruction of a synthetic microstructure from its descriptor. As an example, a central goal of inverse material design, namely the inversion of structure-property linkages, requires a reconstruction. However, a unified, mathematically sound and rigorous, accurate and efficient computational microstructure reconstruction framework is still a demanding task.

For decades, several reconstruction algorithms have been proposed.
A comprehensive review is provided in~\cite{li_review_2014, bostanabad_computational_2018}.
In the following, only a brief introduction is given, where a distinction is made between classical \textit{stochastic} and novel \textit{machine learning} approaches.

\emph{Stochastic} techniques approach microstructure reconstruction as a stochastic optimization problem, iterating on an intermediate microstructure until the corresponding descriptor reaches the desired value. The foundation for these techniques was laid by the Yeong-Torquato algorithm~\cite{yeong1998}, a restriction of the stochastic optimizer ``simulated annealing'' to a search space of microstructures with correct phase volume fractions. 
This versatility has led to Yeong-Torquato variants based on $n$-point correlation functions~\cite{jiao_modeling_2007}, physical descriptors~\cite{xu_descriptor-based_2014}, Gaussian random fields~\cite{jiang2013} and more, allowing for the synthetic generation of a broad spectrum of heterogeneous media and attaining impressive results, see~\cite{bostanabad_computational_2018} and references therein. 
The critical issue with the Yeong-Torquato algorithm is the computational cost, which scales unfavorably with the microstructure resolution and the desired accuracy. To that end, numerous contributions have been made~\cite{bostanabad_computational_2018}, ranging from superior descriptors~\cite{jiao_superior_2009, jiang2013, talukdar2002, li_review_2014, alexander2009} and only partially recomputing them in each step~\cite{yeong_reconstructing_1998, cule_generating_1999} to suitable acceptance distributions~\cite{dueck_threshold_1990, dueck_new_1993, rozman_efficient_2001}, sophisticated cool-down schedules~\cite{li_review_2014} and intelligent ways to propose pixel swaps~\cite{rozman_efficient_2001, zhao_new_2007, jiao_modeling_2008, tang_pixel_2009, pant_stochastic_2014}. 
Enormous progress has been made, nevertheless, the computational cost still forms an obstacle to overcome for broader engineering application.

\textit{Machine learning} has inspired researchers to solve the problem of computational cost through a paradigm shift towards data-driven approaches. The proposed solutions come in great variety, ranging from non-parametric resampling~\cite{bostanabad_stochastic_2016, bostanabad_characterization_2016} and convolutional neural networks (CNNs)~\cite{li_transfer_2018, bostanabad_reconstruction_2020} to variational autoencoders~\cite{cang_microstructure_2017} and generative adversarial networks~\cite{yang_microstructural_2018, fokina_microstructure_2020}. Nevertheless, they are all characterized by an initial \emph{learning} or \emph{training} phase and a concomitant need for a potentially large training data set. In light of the prohibitively high effort needed to generate true microstructure data, transfer learning becomes promising, where the training phase is not omitted, but carried out on a set of different but similar data. This technique has proven its benefits in deep CNN-based synthesis, where the backpropagation gradient based on internal activations of a pre-trained network is diverted to train its input~\cite{li_transfer_2018, bostanabad_reconstruction_2020}. 
Machine learning approaches are particularly attractive due to their efficiency, flexibility and applicability to generic materials.
However, they share a common shortcoming in that the descriptor is not prescribed by the user, but learned from data during (pre-) training in terms of an internal latent representation, which can neither be prescribed nor directly interpreted. For example, machine learning approaches do not yet allow the reconstruction from a given value of the two-point correlation, but only from given coordinates in their internal latent space. This inherent coupling between the (pre-) training data set, the resultant latent space replacing the descriptor and the reconstruction algorithm constitutes an impediment. 

The objective of this paper is to propose a generalized framework for reconstructing random heterogeneous media based on explicitly prescribed descriptors via higher-order optimization under consideration of the gradient. The central idea is the introduction of differentiable microstructure characterization and reconstruction (differentiable MCR), requiring differentiable microstructure descriptors. As an example for suitable differentiable descriptors, we formulate a conceptually simple and robust method based on spatial correlations.
In order to deepen the understanding of this framework, its conceptual similarities with the existing \textit{stochastic} and \textit{machine learning} approaches are discussed. In this regard, it is especially emphasized that deep CNN synthesis as presented in~\cite{li_transfer_2018} emerges as a special form of differentiable MCR. The proposed approach yields accurate and scalable results, reaching a correlation error of $0\%$ in short time. With respect to the Yeong-Torquato algorithm, it constitutes a significant reduction in computational cost, several orders of magnitude, depending on the required accuracy. 
After this introduction, Section~\ref{sec:Yeong-Torquato algorithm} presents the main features of the standard Yeong-Torquato algorithm, which serves as a basis for the proposed concept. Section~\ref{sec:theory} introduces the differentiable MCR approach, whose similarities to existing approaches are discussed in Section~\ref{sec:generalizations}. Section~\ref{sec:exp} presents several example reconstructions and compares the performance to the Yeong-Torquato algorithm. Conclusions are drawn in Section~\ref{sec:summary}. 

The following notation is used in this work: Both upper and lower case letters ($i, \, I$) indicate scalars, $(\vec{r}, \vec{R})$ denote vectors and bold letters ($\boldsymbol{m}, \boldsymbol{M}$) represent arrays with scalar entries ($m_{ij}, M_{ij}$). The implicit summation convention is not used. $\mathbb{R}^{I\times J}$ is an open bounded domain, representing an ${I\times J}$-dimensional array.
This work is restricted to two-dimensional two-phase microstructures.

\section{Revisiting the Yeong-Torquato algorithm}
\label{sec:Yeong-Torquato algorithm}
The reconstruction of random heterogeneous media based on the seminal idea of Yeong and Torquato~\cite{yeong1998} dates back to the late ’90s. Since then, several advancements have been proposed and the literature on this topic is broad, see~\cite{li_review_2014, bostanabad_computational_2018} for a comprehensive review. The most standard form of the algorithm is presented in this section. 
 
The Yeong-Torquato algorithm recasts microstructure reconstruction as an stochastic optimization problem 
\begin{equation}
\underset{\boldsymbol{{M}} \in \{0, \, 1\}^{I \times J}}{\mathrm{argmin}} \; \mathcal{L}(\boldsymbol{{M}}) \, 
\label{YT-Min_Prob}
\end{equation}
where~$\boldsymbol{{M}}$ is a discrete two-phase microstructure of width~$I$ and height~$J$ and can be thought of as an image with $I \times J$ pixels identified by the indices~$i = 1 .. I$ and~$j = 1 .. J$, each carrying its phase number as value, i.e.~0 or~1.
The loss function $\mathcal{L}$ in~(\ref{YT-Min_Prob}) is defined as an error measure in the Euclidean distance ($L_2$) between the current microstructure descriptor $\boldsymbol{{D}}$ and the desired descriptor $\boldsymbol{D}^{\text{des}}$ and is given by
\begin{equation}
	\mathcal{L}(\boldsymbol{{M}}) =  || \boldsymbol{{D}} - \boldsymbol{D}^{\text{des}} ||_2 \; 
\label{L_func_YT}
\end{equation}
where $\boldsymbol{D}$ is obtained from a suitable characterization function
\begin{equation}
\boldsymbol{f_C} : \boldsymbol{M} \in \{0, \, 1\}^{I \times J} \to \boldsymbol{D} \in \mathbb{R}^K
\label{YT-fc}
\end{equation}
which maps \emph{only from discrete-valued} microstructures to the corresponding descriptors~$\boldsymbol{D}$. Herein, $K$ is independent from $I$ and $J$ in the general case. In~(\ref{L_func_YT}), both the characterization function~$\boldsymbol{f_C}$ and its value, the descriptor~$\boldsymbol{D}$, can be chosen independently, allowing to prescribe any descriptor, for example $n$-point correlations, physical descriptors or other metrics. 

In the stochastic optimization procedure, to avoid possible local minima, a simulated annealing optimizer is used to solve~(\ref{YT-Min_Prob}). The pseudo-code for what the authors consider the most basic variant~\cite{torquato_statistical_2002} is presented in Algorithm~\ref{alg:yt}. According to Algorithm~\ref{alg:yt}, the optimizer does not move freely through the space of possible microstructures. Instead, $\boldsymbol{M}$ is initialized with correct volume fractions and in each step, the optimizer swaps two pixel values. It includes many tuneable parameters, such as the acceptance probability distribution function (pdf)~$p$, the cooldown factor~$\alpha \in (0, \, 1)$, the initial temperature\footnote{The temperature~$\vartheta$ regulates the degree of stochasticity of the optimizer through the acceptance probability distribution function. It is decreased in each iteration and motivates the term ``annealing'', as the temperature plays a similar role in physical annealing processes.}~$\vartheta_0$ and, most importantly, as mentioned, a flexible microstructure characterization function~$\boldsymbol{f_C}$. The latter aspect decouples the reconstruction procedure from the chosen characterization metric.

 \begin{algorithm}[htb]
\DontPrintSemicolon
\SetAlgoLined
\KwIn{characterization function $\boldsymbol{f_C}$; desired descriptor $\boldsymbol{D}^\text{des}$; desired volume fractions $v_f$; tolerance $\varepsilon$; initial temperature $\vartheta_0$; cooldown factor $\alpha$; acceptance distribution pdf $p$}
 $\boldsymbol{M} \gets $ random initialization using $v_f$ \;
 $\boldsymbol{D} \gets \boldsymbol{f_C}(\boldsymbol{M})$ \tcp*{initial value for descriptor} 
 $\vartheta \gets \vartheta_0$ \;
 \While(\tcp*[f]{iterate until tolerance reached}){$|| \boldsymbol{D}^\text{des} - \boldsymbol{D} ||_2 > \varepsilon$}{
  $\boldsymbol{M}' \gets$ mutation of $\boldsymbol{M}$ \tcp*{propose to swap two pixels of different color} 
  $\boldsymbol{D}' \gets \boldsymbol{f_C}(\boldsymbol{M}')$ \tcp*{descriptor value for mutated microstructure} 
  $\Delta \mathcal{L} \gets || \boldsymbol{D}^\text{des} - \boldsymbol{D}' ||_2 - || \boldsymbol{D}^\text{des} - \boldsymbol{D} ||_2 $ \tcp*{change in descriptor error}
  \uIf(\tcp*[f]{if mutation has not impaired solution}){$\Delta \mathcal{L} \leq 0$}{
   $\boldsymbol{M}, \, \boldsymbol{D} \gets \boldsymbol{M}', \, \boldsymbol{D}'$ \tcp*{apply mutation}
   }
   \ElseIf(\tcp*[f]{if mutation has worsened solution}){$p(\Delta \mathcal{L}, \vartheta) < r \sim \mathcal{U}(0,1)$}{
   $\boldsymbol{M}, \, \boldsymbol{D} \gets \boldsymbol{M}', \, \boldsymbol{D}'$ \tcp*{apply mutation only with certain acceptance probability}
   }
  $\vartheta \gets \alpha \cdot \vartheta$ \tcp*{decrease temperature}
 }
\KwOut{Microstructure $\boldsymbol{M}$}
 \caption{Yeong-Torquato algorithm:  Minimization of $\mathcal{L}$ via simulated annealing method.}
 \label{alg:yt}
\end{algorithm}
 
The crucial aspect of the Yeong-Torquato algorithm is performance. Consider a microstructure $\boldsymbol{M}$ with $I = 100$ and $J = 100$ pixels.
With the volume fraction $v_f = 50\%$ of both phases, the number of possible microstructures is given by the binomial coefficient $\binom{I \cdot J}{v_f \cdot (I \cdot J)} = \binom{10000}{5000} \approx 1.6 \cdot10^{3008}$.
This is why the Yeong-Torquato algorithm searches the space of possible microstructures not by brute force, but by using simulated annealing. However, in light of the sheer number of possibilities and the inherent randomness of the heuristic, the efficiency of the basic Yeong-Torquato algorithm is naturally insufficient. 
To that end, as already mentioned in Section~\ref{sec:intro}, numerous contributions have been made, ranging from superior descriptors and only partially recomputing them in each step to suitable acceptance distributions, sophisticated cool-down schedules and intelligent ways to propose pixel swaps~\cite{bostanabad_computational_2018}. The performance gains are considerable, yet, despite highly optimized implementations and ever better heuristics, the procedure essentially remains random. This randomness affects the convergence behavior, especially in the later stages, since random mutations are increasingly unlikely to lower $\mathcal{L}$ as the microstructure evolves. This is further discussed based on the numerical examples in Section~\ref{sec:exp}.

The alternative to randomly guessing each step's microstructure mutation is to infer the optimal mutation from $\mathcal{L}$ by computing and descending its gradient. The difficulties thereby lie in the formulation of a differentiable characterization function and in avoiding local minima. Despite these challenges, the potential performance gains motivate the main hypothesis of this contribution: Microstructure reconstruction is optimization and efficient optimization requires the gradient.

\section{Central concept of differentiable MCR}
\label{sec:theory}
In this section, we first introduce the reconstruction framework for random heterogeneous media based on differentiable optimization. With this concept at hand, we develop the mathematical formulation of differentiable MCR and exemplarily apply its generic descriptor concept to spatial correlations. The key difference between the Yeong-Torquato algorithm and the suggested one lies in providing the gradient information to the optimizer, which, in turn, requires a differentiable microstructure $\boldsymbol{\tilde{M}}$ and its descriptor~$\boldsymbol{\tilde{D}}$. The subsequent formulation is presented in the following Sections~\ref{sec:formulation} and~\ref{sec:corrs}. 

\begin{figure}[ht]
    \centering
    \includegraphics[width = \linewidth]{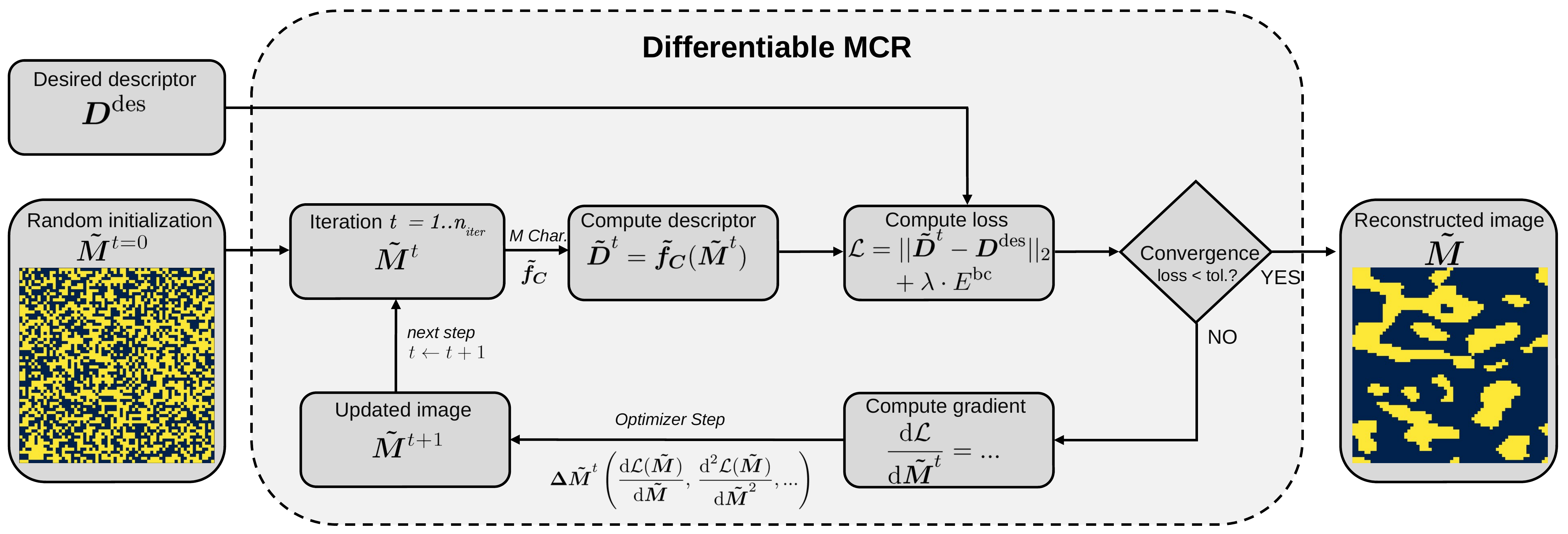}
    \caption{Flow chart illustrating the proposed differentiable microstructure characterization and reconstruction (MCR).}
    \label{fig:flowchart}
\end{figure}

The iterative procedure to minimize the reconstruction error is illustrated in Figure~\ref{fig:flowchart}. Starting from a randomly initialized microstructure~$\boldsymbol{\tilde{M}}^{0}$ at iteration $t=0$, a generic microstructure characterization function~$\boldsymbol{\tilde{f}_C}$ yields the corresponding current descriptor~$\boldsymbol{\tilde{D}}^t$, which may be freely chosen as a classical descriptor, a spatial correlation, or any other suitable metric. The current value is compared to the desired one through the loss function $\mathcal{L}$, which contains the Euclidean distance between the descriptors as well as additional penalty terms for the boundary conditions of the optimization problem. If the loss is sufficiently small, the algorithm has converged, as the reconstructed medium complies with (almost) all of the prescribed characteristics. Otherwise the current microstructure $\boldsymbol{\tilde{M}}^t$ is altered by an optimizer to decrease the loss and the next iteration begins. 

Here, the core idea lies in how $\boldsymbol{\tilde{M}}^{t+1}$ is computed. In differentiable MCR, the intermediate microstructure is updated considering derivatives of the loss function with respect to the microstructure
\begin{equation}
    \boldsymbol{\tilde{M}}^{t+1} \gets \boldsymbol{\tilde{M}}^{t} + \boldsymbol{\Delta \tilde{M}}^t \left( \dfrac{\text{d} \mathcal{L} (\boldsymbol{\tilde{M}}^{}) }{\text{d} \boldsymbol{\tilde{M}}^{}}, \, \dfrac{\text{d}^2 \mathcal{L} (\boldsymbol{\tilde{M}}^{}) }{\text{d} \boldsymbol{\tilde{M}}^{2}}, ... \right).    
\label{M-update}
\end{equation}
where the update $\boldsymbol{\Delta \tilde{M}}^t$ is applied to the current microstructure $\boldsymbol{\tilde{M}}^t$ of step $t$. Consequently, it becomes necessary to formally redefine microstructure reconstruction as a differentiable optimization problem and to reformulate the properties of microstructure descriptors such that they are differentiable. This differentiability of the descriptors allows to compute the gradients by evaluating an (exact) analytical representation, not just through numerical approximation. The specific procedure to compute $\boldsymbol{\Delta \tilde{M}}^t$ depends on the chosen optimizer. In this work, only first-order gradient terms are considered in Equation~(\ref{M-update}), although higher-order derivatives can be included.

\subsection{Formulation of differentiable MCR}
\label{sec:formulation}
In this contribution, we propose to reformulate~(\ref{YT-Min_Prob}) as differentiable optimization problem
\begin{equation}
\underset{\boldsymbol{\tilde{M}} \in \mathbb{R}^{I \times J}}{\mathrm{argmin}} \; \mathcal{L}(\boldsymbol{\tilde{M}}) \; 
\label{dMCR-Min_Prob}
\end{equation}
where each pixel of the intermediate microstructure $\boldsymbol{\tilde{M}} \in \mathbb{R}^{I \times J}$ must be allowed to take any real value\footnote{Also values outside the interval $(0, \, 1)$ are allowed during iteration. Later, a penalty term is introduced to enforce the interval as a boundary condition, but nevertheless, $\boldsymbol{\tilde{f}_C}$ must be defined for such cases.} during the reconstruction process, unlike ${\boldsymbol{{M}} \in \{0, \, 1\}^{I \times J}}$ in ~(\ref{YT-Min_Prob}). We further demand that the minimizer should use the gradient of the cost function with respect to the microstructure ${\text{d} \mathcal{L}} / {\text{d} \boldsymbol{\tilde{M}}}$ while updating the intermediate microstructure $\boldsymbol{\tilde{M}}$ in~(\ref{M-update}). 

For this purpose, the loss function in~(\ref{dMCR-Min_Prob}) is defined as
\begin{equation}
	\mathcal{L}(\boldsymbol{\tilde{M}}) =  E(\boldsymbol{\tilde{M}}) + \lambda(\boldsymbol{\tilde{M}}) \cdot E^{\text{bc}}(\boldsymbol{\tilde{M}}) \; 
	\label{L_func_dMCR}
\end{equation}
where $E(\boldsymbol{\tilde{M}})$ is given by
\begin{equation}
E(\boldsymbol{\tilde{M}}) = || \boldsymbol{\tilde{D}} - \boldsymbol{D}^{\text{des}} ||_2 \; .
\end{equation}
To obtain $\boldsymbol{\tilde{D}} = \boldsymbol{\tilde{f}_C}(\boldsymbol{\tilde{M}})$, the concept of~(\ref{YT-fc}) must be generalized for differentiable microstructure reconstruction as $\boldsymbol{\tilde{M}}$ can take any real value. This requires a characterization function
\begin{equation}
\boldsymbol{\tilde{f}_C} : \boldsymbol{\tilde{M}} \in \mathbb{R}^{I \times J} \to \boldsymbol{\tilde{D}} \in \mathbb{R}^K
\label{eqn:fctilde}
\end{equation}
that takes a wider range of arguments than $\boldsymbol{f_C}$.
In the special case of an integer-valued microstructure $\boldsymbol{M}$, $\boldsymbol{\tilde{f}_C}$ should be indistinguishable from $\boldsymbol{f_C}$, such that
\begin{equation}
\boldsymbol{\tilde{f}_C}(\boldsymbol{M} \in \{0, \, 1\}^{I \times J}) = \boldsymbol{D} \, .
\end{equation}
In~(\ref{L_func_dMCR}), $E^{\text{bc}}(\boldsymbol{\tilde{M}})$ is introduced to enforce the boundary condition
\begin{equation}
0 \leq \tilde{M}_{ij} \leq 1 \quad \forall i \in 1..I \quad \forall j \in 1..J
\end{equation}
via
\begin{equation}
E^{\text{bc}}(\boldsymbol{\tilde{M}}) = \sum_{i, \, j} \left[ \dfrac{(\tilde{M}_{ij}-1) \, H(\tilde{M}_{ij}-1)}{I \cdot J} + \dfrac{-\tilde{M}_{ij} \, H(-\tilde{M}_{ij})}{I \cdot J} \right]^2\; 
\end{equation}
with $H(x)$ as a Heaviside-function 
\begin{equation}
H(x) = \left\{
\begin{array}{ll}
0, & x \leq 0 \\
1, & x > 0 \\
\end{array}
\right. \\
\label{eqn:heavy}
\end{equation}
and ${\lambda}(\boldsymbol{\tilde{M}})$ as a scalar weight to the boundary condition. The weight ${\lambda}$ can be fixed to a constant, but since the boundary condition should neither dominate the cost function nor vanish at any stage of convergence, it is set as proportional to the reconstruction error term
\begin{equation}
\lambda(\boldsymbol{\tilde{M}}) \propto E(\boldsymbol{\tilde{M}}).
\end{equation}

To solve the optimization problem~(\ref{dMCR-Min_Prob}) considering the gradient term ${\text{d} \mathcal{L}} / {\text{d} \boldsymbol{\tilde{M}}}$ in~(\ref{M-update}) necessitates additional requirement for the characterization function~$\boldsymbol{\tilde{f}_C}$. It is evident that $\boldsymbol{\tilde{f}_C}$ should be \emph{continuously differentiable}. 
In addition to this,~$\boldsymbol{\tilde{f}_C}$ has to satisfy further requirements because gradient-based optimizers are prominently susceptible to terminating in a local minimum or on a plateau where the gradient is zero. 
The latter issue can be addressed by using suitable minimizers, for instance with momentum~\cite{kingma_adam_2017} and by defining an appropriate characterization function $\boldsymbol{\tilde{f}_C}$ such that its gradient is non-zero for the widest possible range of arguments, ideally for all inputs (see Section~\ref{sec:corrs}). The former issue is more subtle. Local minima are the reason why the Yeong-Torquato algorithm was formulated as a stochastic procedure~\cite{torquato_statistical_2002}. However, in the Yeong-Torquato algorithm, where only microstructures in $\{0, \, 1\}^{I \times J}$ are allowed, the optimizer is restricted to swapping pairs of pixels. If, in contrast, it was allowed to swap any number of pixels at the same time, then there would always be a vanishingly small but non-zero probability of directly reaching the global minimum in the next step. Thus, local minima impede the Yeong-Torquato algorithm precisely \emph{because} the optimizer is restricted to swapping pairs of pixels. Conversely, in differentiable MCR, the optimizer may move freely in $\mathbb{R}^{I \times J}$. The existence or absence of local minima hence does not depend on $\boldsymbol{f_C}$, but only on how it is generalized to $\boldsymbol{\tilde{f}_C}$. This is further discussed in Section~\ref{sec:li} with empirical evidence.

\subsection{Application of differentiable MCR to spatial correlations} 
\label{sec:corrs}
While the considerations in Section~\ref{sec:formulation} apply to any descriptor, we now specifically apply the process of replacing the characterization function $\boldsymbol{f_C}$ (\ref{YT-fc}) by its differentiable generalization $\boldsymbol{\tilde{f}_C}$~(\ref{eqn:fctilde}) to spatial correlations. 
First, a brief introduction to spatial correlations is given. Then, although there are numerous highly-efficient ways of computing spatial correlations, ranging from Fourier-based approaches~\cite{brough_materials_2017} to algorithms that benefit from only partially recomputing the correlations after a microstructure update~\cite{yeong_reconstructing_1998}, we introduce a conceptually simple and robust method which is then cast into a differentiable function suitable for the proposed differentiable MCR approach.

\subsubsection{Brief introduction to spatial correlations}
\label{sec:corrsintro}
Amongst all available microstructure descriptors, spatial $n$-point correlation functions ($n \geq 2$) belong to the most popular ones due to their simplicity and generality.  A detailed definition and discussion can be found in~\cite{jiao_modeling_2007}. In the following, a brief description is provided.

\begin{figure}[ht]
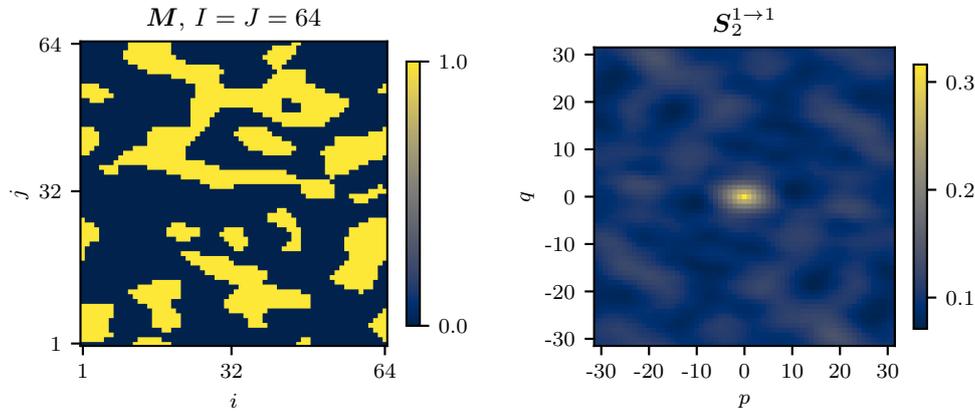

    \centering
    \input{example_ms.pgf}
    \input{example_correlation.pgf}
    \caption{An exemplary microstructure generated synthetically via \emph{pyMKS}~\cite{brough_materials_2017} and its corresponding descriptor. Left: example microstructure~$\boldsymbol{M}$; right: corresponding two-point correlation~$\boldsymbol{S}_2^{1 \to 1}\left(\vec{r} \right)$ over the vector components~$p$ and~$q$. }
    \label{fig:examplecorrelation}
\end{figure}

Given a vector~$\vec{r}$ with components~$p$ and~$q$, the two-point correlation~$\boldsymbol{S}_2^{a \to b}(\vec{r})$ yields the probability of~$\vec{r}$ starting in phase~$a$ and ending in phase~$b$ when~$\vec{r}$ is placed randomly in the microstructure. If $a = b$, one speaks of auto-correlations, otherwise of cross-correlations. In a two-phase microstructure with phases~$0$ and~$1$, it suffices to compute a single auto-correlation, for example~$\boldsymbol{S}_2^{1 \to 1}$, as~$\boldsymbol{S}_2^{0 \to 0}$, $\boldsymbol{S}_2^{0 \to 1}$ and~$\boldsymbol{S}_2^{1 \to 0}$ can be derived from it. For any number of phases, the set of non-redundant correlations is described in~\cite{niezgoda_delineation_2008}. An exemplary illustration of~$\boldsymbol{S}_2^{1 \to 1}$ is given in Figure~\ref{fig:examplecorrelation}.

Generalizing this concept to $n-1$ vectors yields the complete set of $n$-point correlations, which fully describes a microstructure if $n$ is sufficiently large. The high-order correlations ($n \geq 3$), however, become too much data to compute and store in practice. In this work, besides $\boldsymbol{S}_2^{1 \to 1}(\vec{r})$, we restrict ourselves to the three-point auto-correlation of phase 1 $\boldsymbol{S}_3^{1 \to 1}(\vec{r}_a, \, \vec{r}_b)$, which maps the vectors $\vec{r}_a$ and $\vec{r}_b$ to the probability of both vectors starting and ending in phase 1, if their mutual starting point is placed randomly in the microstructure.

\subsubsection{Computing correlations from ensembles}
\label{sec:e2s}
As vectors are placed randomly in the microstructure, it is reasonable to compute a statistical ensemble for a given microstructure~$\boldsymbol{M}$ and vector~$\vec{r}$, containing realizations of~$\vec{r}$ randomly placed in~$\boldsymbol{M}$. Each realization in the ensemble takes the value~1 if both ends are in phase~1, otherwise it is~0, as illustrated in Figure~\ref{fig:fc}. As the number of realizations $n_r$ in the ensemble grows, the ensemble average converges to the two-point correlation. As the microstructure is discretized into a finite set of pixels, the need to generate realizations randomly and letting~$n_r \to \infty$ can be avoided. It is sufficient to consider each pixel in~$\boldsymbol{M}$ as a starting point for~$\vec{r}$, so that $n_r = I \cdot J$~\cite{yeong_reconstructing_1998}. 
For multiple different vectors, we define the ensemble~$\boldsymbol{e}$, where each entry~$e_{ijpq}$ stands for the realization corresponding to placing $\vec{r}= \left( p \quad q \right)^{\mathrm T}$ in the microstructure, with microstructure entry $M_{ij}$ as a starting point. 

A linear ensemble average of $\boldsymbol{e}$ yields the two-point autocorrelation
\begin{equation}
	S_{pq}^{1 \to 1} = \boldsymbol{S}_2^{1 \to 1}\left( \vec{r} = \begin{pmatrix} p \\ q \end{pmatrix} \right) = \sum_{i, \, j} \dfrac{e_{ijpq}}{I \cdot J} \; .
\end{equation}
Likewise, a quadratic reduction yields the two- and three-point correlations
\begin{equation}
S_{pqrs}^{1 \to 1} = \boldsymbol{S}_3^{1 \to 1}\left(\begin{pmatrix} p \\ q \end{pmatrix} , \; \begin{pmatrix} r \\ s \end{pmatrix}  \right) = \sum_{i, \, j} \dfrac{e_{ijpq} e_{ijrs}}{I \cdot J} \; 
\label{eqn:s3}
\end{equation}
where each element $S_{pqrs}^{1 \to 1}$ represents a two-point correlation if $p = r$ and $q = s$ and a three-point correlation otherwise. 
Following the same archetype, higher-order correlations can be obtained through appending more ensembles to~(\ref{eqn:s3}).
With this strategy of obtaining correlations from ensembles, the computation of the ensembles is discussed in the next Section~\ref{sec:m2e}.

\subsubsection{Computing ensembles through convolutions}
\label{sec:m2e}
\begin{figure}[ht]
    \centering
    \input{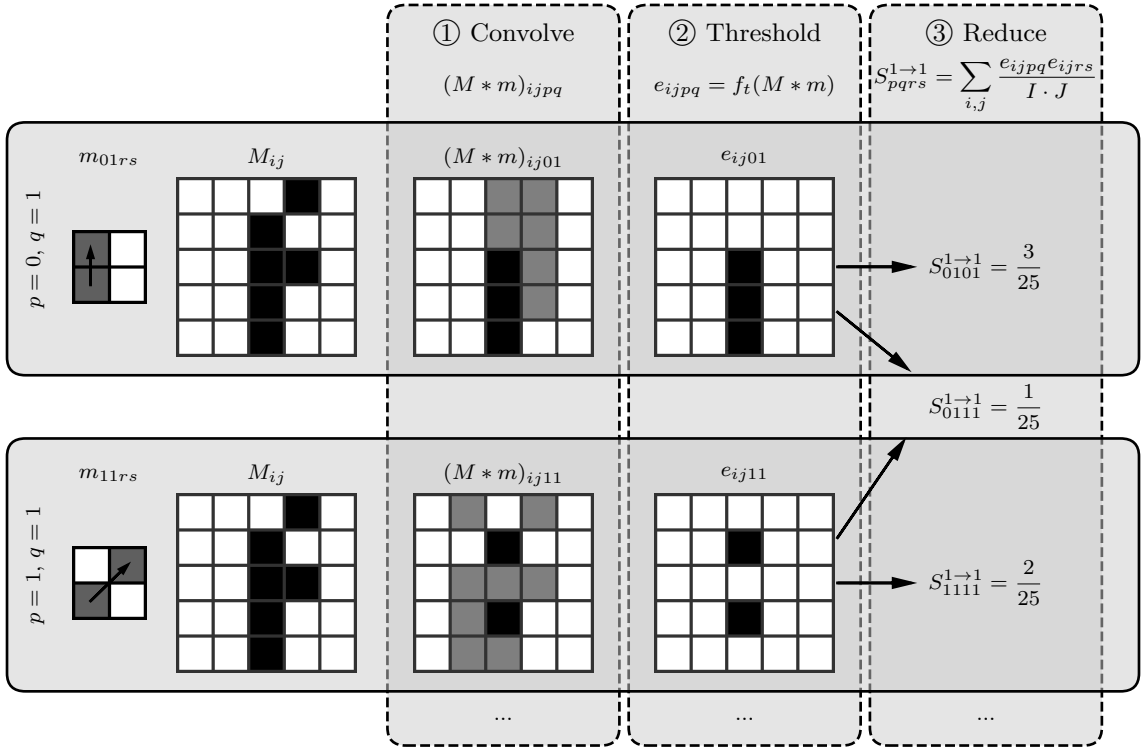}
    \caption{Visualization of the \emph{convolve} - \emph{threshold} - \emph{reduce} pipeline for computing spatial correlations. Black corresponds to the value~1, white to~0 and grey to~\nicefrac{1}{2}.}
    \label{fig:fc}
\end{figure}

We introduce a mask $\boldsymbol{m}$ as an array with entries $m_{pqrs}$, where the first two indices identify the corresponding $\vec{r}$ and the last two span the convolution mask associated with that $\vec{r}$. For instance, a mask corresponding to the $\vec{r}$ for the chosen components $p=0$ and $q=1$ is again a two-dimensional array with entries $m_{01rs}$, where~$r$ and~$s$ identify the column and row of that particular convolution mask respectively.
Generally, such a convolution mask can be obtained by placing $\vec{r}$ on an all-zero initialized $\boldsymbol{m}$ and increasing the value by $\nicefrac{1}{2}$ at the tip and end\footnote{For $\vec{r} \neq \vec{0}$, $\boldsymbol{m}$ thus contains mostly zeros and two entries with value $0.5$. For $\vec{r} = \vec{0}$, $\boldsymbol{m}$ contains mostly zeros and one entry with value $1$} of $\vec{r}$, see Figure~\ref{fig:fc}. 

\textit{Firstly}, in the \emph{convolve} step, $\boldsymbol{M}$ is convolved with this mask $\boldsymbol{m}$, such that
\begin{equation}
	\boldsymbol{M} * \boldsymbol{m} = \left\{
	\begin{array}{ll}
	0, & \text{no end of $\vec{r}$ in phase 1} \\
	\frac{1}{2}, & \text{exactly one end in phase 1} \\
	1, & \text{both ends in phase 1} \\
	\end{array}
	\right. \;  \\
\end{equation}
whereas, in a more general case, with the symbolic convolution operator $*$, it is given by
\begin{equation}
    (\boldsymbol{M} * \boldsymbol{m})_{ijpq} = \sum_{k, l, r, s} M_{kl} \delta^k_{\text{mod}_I(i+r) + 1} \delta^l_{\text{mod}_J(j+s) + 1} m_{pqrs} \; 
    \label{eqn:convdef}
\end{equation}
where $\delta$ denotes the Kronecker-Delta
\begin{equation}
    \delta_i^j = \left\{
	\begin{array}{ll}
	1, & i = j \\
	0, & i \neq j \\
	\end{array}
	\right. \;  \\
\end{equation}
and \emph{mod} denotes the modulo operator.

\textit{Secondly}, in the \emph{threshold} step, to obtain 
\begin{equation}
\boldsymbol{e} = {f}_t(\boldsymbol{{M}} * \boldsymbol{m}) \; 
\end{equation}
from the convolution result, the $\nicefrac{1}{2}$-cases must be dropped to zero by applying a suitable thresholding function~$f_t(x)$, for example
\begin{equation}
	f_t(x) = H(x - 0.75) \; 
	\label{f_t}
\end{equation}
where $H(x)$ denotes the Heaviside step-function defined in~(\ref{eqn:heavy}).

\emph{Thirdly}, in the \emph{reduce} step, as presented in Section~\ref{sec:e2s}, $\boldsymbol{S}^{1\to 1}_3$ can be obtained from $\boldsymbol{e}$ through~(\ref{eqn:s3}). Figure~\ref{fig:fc} illustrates the complete algorithm to compute two- and three-point correlations of integer-valued microstructures using a triplet pipeline of \emph{convolve} - \emph{threshold} - \emph{reduce}. As can be observed, the proposed method yields the exact spatial $n$-point correlation functions for $\boldsymbol{f_C}$. 
In the following Section~\ref{sec:s2stilde}, this method is now extended to $\boldsymbol{\tilde{f}_C}$ to map from a real-valued microstructure $\boldsymbol{\tilde{M}} \in \mathbb{R}^{I \times J}$ to a descriptor $\boldsymbol{\tilde{D}}$, while at the same time being continuously differentiable and recovering $\boldsymbol{D}$ for integer-valued microstructures $\boldsymbol{M}$. 

\subsubsection{Differentiable correlations}
\label{sec:s2stilde}
\begin{figure}[ht]
    \centering
    \input{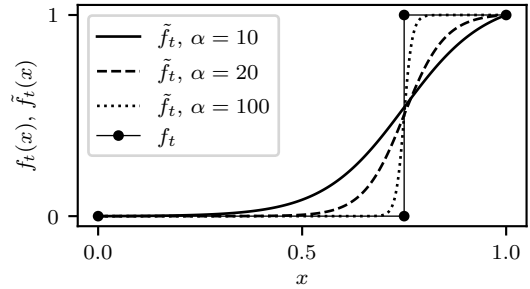}  
    \caption{To make the \emph{convolve} - \emph{threshold} - \emph{reduce} pipeline differentiable, the threshold~$f_t(x)$ is replaced by the soft threshold~$\tilde{f}_t(x)$ with slope parameter~$\alpha$.}
    \label{fig:ft}
\end{figure}

In order to obtain the differentiable characterization function $\boldsymbol{\tilde{f}_C}$ using the \emph{convolve} - \emph{threshold} - \emph{reduce} pipeline described in Section~\ref{sec:m2e}, note that the convolution and the reduction operations are continuously differentiable, whereas only the thresholding function $f_t$ is non-differentiable. For the purpose of allowing it to be a differentiable function, $f_t$ is replaced by a so-called \emph{soft threshold} function $\tilde{f}_t$, see Figure~\ref{fig:ft}. We choose a shifted and scaled version of the widely used logistic function\footnote{NB: The often-used tanh-function can be expressed via setting $c_1=\alpha=2$, $c_2=0$ and $c_3=-1$}
\begin{equation}
    \tilde{f}_t(x) = \dfrac{c_1}{1 + \text{e}^{-\alpha (x + c_2)}} + c_3 \; 
\label{eqn:softthreshold}
\end{equation}
where $c_1$, $c_2$ and $c_3$ are defined by requiring
\begin{equation}
    \tilde{f}_{t}(0) = 0 ; \qquad \tilde{f}_{t}(1) = 1 ; \qquad \tilde{f}_{t}''(0.75) = 0 \, 
\end{equation}
and $\alpha$ is a hyperparameter for the slope of $\tilde{f}_t$ which is chosen as $\alpha = 10$. Note that $\lim_{\alpha \to \infty} \tilde{f}_t = f_t$.

Alternative choices for $\tilde{f}_t$ are equally conceivable, if the requirements discussed in Section~\ref{sec:formulation} are satisfied, namely $\tilde{f}_t$ should not only be continuously differentiable and $\tilde{f}_{t}'(x) > 0 \; \forall x \in (0, 1)$ should also hold. This is the case for the chosen function (\ref{eqn:softthreshold}). 
Furthermore, the latter prerequisite of forcing the gradient to be non-zero prevents the optimizer from deadlocking due to a vanishing gradient, but further implies that $\tilde{f}_t(0.5) > 0$. This disintegrates the motivation of the thresholding function~(\ref{f_t}), which is to drop all $\frac{1}{2}$-values from $\boldsymbol{M} * \boldsymbol{m}$, see Figure~\ref{fig:ft}. Consequently, $\boldsymbol{\tilde{e}} = \boldsymbol{\tilde{f}_t} (\boldsymbol{\tilde{M}} * \boldsymbol{m})$ loses its interpretability as an ensemble. We therefore refer to the quadratic reduction $\frac{1}{I \cdot J} \sum \boldsymbol{\tilde{e}} \boldsymbol{\tilde{e}}$ as $\boldsymbol{\bar{S}}^{1 \to 1}$ instead of $\boldsymbol{\tilde{S}}^{1 \to 1}$. Fortunately, despite different numerical values, $\boldsymbol{\bar{S}}^{1 \to 1}$ contains the same information as $\boldsymbol{\tilde{S}}^{1 \to 1}$, allowing for the definition of a mapping between the two, which is clarified in Appendix~\ref{sec:correction}. If the exact spatial correlations are to be computed, this mapping must be appended to the \emph{convolve} - \emph{threshold} - \emph{reduce} pipeline as a soft threshold correction step. However, as $\boldsymbol{\bar{S}}^{1 \to 1}$ and $\boldsymbol{\tilde{S}}^{1 \to 1}$ contain the same information, the mapping step can be omitted. Such a choice constitutes a change from correlations to a different but equivalent descriptor, hence the numerical values of the desired descriptor $\boldsymbol{D}^{\text{des}}$ must be adapted accordingly.

\subsection{Parallel short-range multigrid descriptors}
\label{sec:mg}
The \emph{convolve} - \emph{threshold} - \emph{reduce} pipeline introduced in Sections~\ref{sec:m2e} and~\ref{sec:s2stilde} contains two undefined parameters, $P$ and $Q$, denoting the bounds of the indices $p, r = -P .. P$ and $q, s = -Q .. Q$ that define the correlation vectors. The natural choice to use $P = \lfloor {I} / {2} \rfloor$ and $Q = \lfloor {J} / {2} \rfloor$ leads to a highly unfavorable time complexity in the convolution step, whereas setting $P$ and $Q$ to a prescribed constant completely disregards all long-range correlations, which reduces the descriptor cost but disproportionately increases the number of iterations required for convergence. We address the issue of obtaining an optimized and suitable descriptor for differentiable MCR by suggesting a technique referred to as \emph{multigrid correlations}. 
\begin{figure}[h]
    \centering
    \input{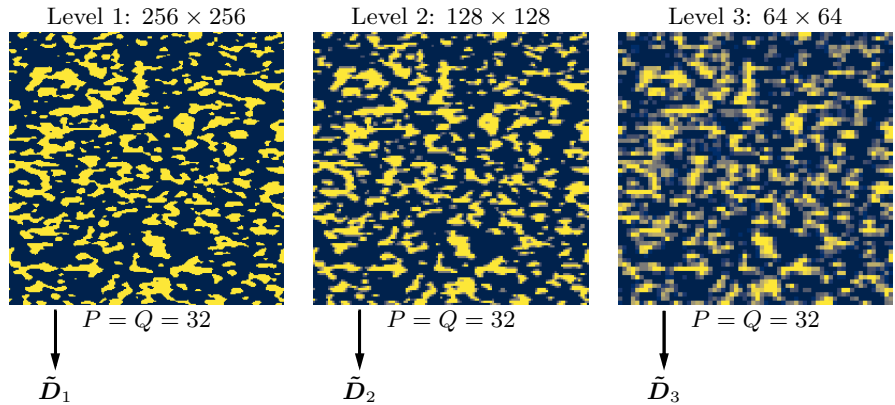}
    \caption{Illustration of multigrid descriptors. Each level contains the same microstructure in a different resolution to compute limited-length descriptors. As the structure coarsens at deeper levels, the descriptor contains longer-range but less accurate information.}
    \label{fig:mg}
\end{figure}

Figure~\ref{fig:mg} illustrates the concept of the multigrid correlations. The multigrid descriptor operates on different levels, each carrying a different resolution representation of the same microstructure~$\boldsymbol{\tilde{M}}$. Each level~$l$ computes the single-grid descriptor $\boldsymbol{\tilde{D}}_l$ with $P$ and $Q$ set to a constant. At the highest image resolution, $\boldsymbol{\tilde{D}}$ contains the high-resolution short-range correlations. As the image resolution decreases, $\boldsymbol{\tilde{D}}$ becomes increasingly coarse whilst long-range correlations are captured. All these different correlations $\boldsymbol{\tilde{D}}_l$ are then weighted and concatenated into a single descriptor 
\begin{equation}
    \boldsymbol{\tilde{D}} = \left[ w_1 \cdot \boldsymbol{\tilde{D}}_1, \, w_2 \cdot \boldsymbol{\tilde{D}}_2, \, .. \, , w_l \cdot \boldsymbol{\tilde{D}}_l \right]
\label{eqn:mg}
\end{equation}
where $w_l$ denotes the weight of layer $l$, which is chosen\footnote{NB: In 2D, layer~$l$ summarizes~$2^{2(l-1)}$ pixels in a single one, creating an imbalance that can be compensated through choosing the layer weights as~$w_l = 2^{2(l-1)}$. However, to account for the higher importance of short-range correlations, the square root of this term is chosen in the present work. Other choices are equally conceivable.} as $w_l = 2^{l-1}$. This technique of computing $\boldsymbol{\tilde{D}}$ is directly applied to correlations presented in Section~\ref{sec:corrs} and used herein, leading a favorable time complexity without having completely omitted the long-range correlations. Unlike previous multigrid methods, which sequentially converge on each layer~\cite{alexander2009}, differentiable MCR allows to operate on all levels simultaneously and the gradient can be backpropagated through all multigrid levels simultaneously. Such a multigrid technique might be equally applicable to other descriptors.

\section{Conceptual similarities to existing approaches}
\label{sec:generalizations}
Although \textit{stochastic} MCR has been a topic of very intense research in the last two decades, the incorporation of gradient information during reconstruction has been the subject of only one publication~\cite{fullwood_gradient-based_2008}. Furthermore, a novel \textit{machine learning} approach relies on it~\cite{li_transfer_2018}. In this section, to deepen the understanding of the new differential MCR approach, we outline its conceptual similarities to previous methods. The original symbols and notations are adapted to the present notation for unification purposes.

\subsection{Stochastic approach}
\label{sec:vollholz}
Amongst the \textit{stochastic} reconstruction approaches, to the author's best knowledge, the only comparable approach in the literature is given in the work of Fullwood~et~al.~\cite{fullwood_gradient-based_2008}.
Herein emphasis is placed on two aspects. First, to compute two-point correlations, a highly efficient method using fast Fourier transforms~(FFTs) is proposed. The FFT-method is defined for real-valued microstructures~$\boldsymbol{\tilde{M}}$. It computes not $\boldsymbol{S}_2$, but $\boldsymbol{\tilde{S}}_2$, although a different version than presented in Section~\ref{sec:corrs}. 
Secondly, the microstructures are reconstructed via the gradient-based optimizer ``sequential quadratic programming'' (SQP) provided by MATLAB 2006~\cite{MATLAB}. The loss function minimized in~\cite{fullwood_gradient-based_2008} is
\begin{equation}
    \mathcal{L}(\boldsymbol{\tilde{M}}^t) = \sum_{k} \sum_{p, q} \omega_k \left[ \mathfrak{F}({\tilde{S}}_{p q}(\boldsymbol{\tilde{M}}^t)) - \mathfrak{F}({\tilde{S}}_{p q}^{\text{des}}) \right]^2
\label{Loss-FFT}
\end{equation}
where $\mathfrak{F}$ denotes the Fourier transform, $k$ the wave number and $\omega_k$ is an unspecified weighting term. 
In~(\ref{Loss-FFT}), $\mathfrak{F}(\tilde{\boldsymbol{S}})$ is not computed explicitly, but $\boldsymbol{S}$ is directly computed in Fourier space. 

With the information available in~\cite{fullwood_gradient-based_2008},  the central aspect whether the optimizer needs to evaluate the gradient of~(\ref{Loss-FFT}) numerically or analytically is unclear. This computation of ${\text{d} \mathcal{L}} / {\text{d} \boldsymbol{\tilde{M}}}$ while updating $\boldsymbol{\tilde{M}}$ in~(\ref{M-update}) constitutes a significant difference in high-dimensional spaces and with non-convex loss functions. In MATLAB, SQP computes the gradient numerically if it is not provided explicitly. Furthermore, a Yeong-Torquato algorithm is implemented based on the Fourier-based descriptor. However, the computational cost of the gradient-based optimizer is reported to be comparable to the Yeong-Torquato algorithm~\cite{fullwood_gradient-based_2008}. 

In the present work, the exact gradient of loss function~(\ref{L_func_dMCR}) is known to the optimizer in every step. The proposed concept of differentiable MCR and its application to spatial correlations in Sections~\ref{sec:formulation} and ~\ref{sec:corrs} rigorously allows the analytical representation of ${\text{d} \mathcal{L}} / {\text{d} \boldsymbol{\tilde{M}}}$. This and possibly many other differences result in a significant reduction of the computational cost with respect to the available standard algorithm, as will be shown in the numerical examples in Section~\ref{sec:exp}. 

\subsection{Machine learning approach}
\label{sec:li}
The present approach can also be motivated from a novel \textit{machine learning} approach of reconstructing random heterogeneous media. We compare it to a seemingly unrelated approach of characterizing and reconstructing microstructures using deep CNNs, specfically, pre-trained deep CNNs that have gained popularity to a variety of image processing tasks.

It is shown by Lubbers~et~al.~\cite{lubbers_inferring_2017} that the VGG-19~\cite{simonyan_very_2015} network's inner layer activations are indeed a powerful microstructure descriptor, although the network is trained exclusively on the imagenet dataset~\cite{russakovsky_imagenet_2015}, which contains no microstructure data at all. This descriptor information is exploited in the work of Li~et~al.~\cite{li_transfer_2018}. Based on a general-purpose texture synthesis approach of Gatys~et~al.~\cite{gatys_texture_2015}, a highly efficient algorithm is presented to reconstruct a microstructure. Starting from a randomly initialized microstructure $\boldsymbol{\tilde{M}}$, the loss function to minimize is defined as the norm of the difference between the Gram\footnote{The Gram matrix~$G_{\bar{p} \bar{r}}^{\bar{n}} = \sum_{\bar{i}} F_{\bar{i}\bar{p}}^{\bar{n}} F_{\bar{i}\bar{r}}^{\bar{n}}$ is a translation-invariant descriptor based on the neural activations of a CNN~\cite{li_transfer_2018}. Herein, to state simply, the activation~$F_{\bar{i}\bar{p}}^{\bar{n}}$ is the value that neuron~$\bar{p}$ of layer~${\bar{n}}$ takes at the spatial position~${\bar{i}}$. The analogies reach deep into the machine learning terminology, but the following similarity between the Gram matrix and the quadratic reduction~(\ref{eqn:s3}) used to obtain $\boldsymbol{S}^{1 \to 1}$ from $\boldsymbol{e}$ may serve as an intuition: Interpreting the ensemble as a set of activations, it can be seen that the two- and three- point correlations are nothing but the Gram matrix associated with the special set of activations herein called ensemble.} matrices $\boldsymbol{G}^{l \text{, des}}$ and $\boldsymbol{G}^{l}$ of the desired and current activations, respectively. It reads as
\begin{equation}
    \mathcal{L}(\boldsymbol{\tilde{M}}^t) = \sum_l \sum_{m, n} \frac{1}{N_l} \left[ {G}^l_{m n}(\boldsymbol{\tilde{M}}^t) - {G}^{l \text{, des}}_{m n} \right]^2 \; 
\label{Loss-CNN}
\end{equation}
where $N_l$ is the normalization of layer $l$ and $\boldsymbol{G}^l$ is the Gram matrix of the inner layer activations of a pre-trained VGG network. 

The common backpropagation algorithm of machine learning is used to compute the gradient of the loss function from~(\ref{Loss-CNN}). However, instead of applying this gradient information to the weights of the network as it is usually done in machine learning, it is applied to refine $\boldsymbol{\tilde{M}}$ in~(\ref{M-update}) iteratively until the loss function is sufficiently small. 
Although the concept of~\cite{li_transfer_2018} is to motivate transfer learning for microstructure reconstruction, as can be observed, the problem is approached as an optimization scheme. The gradient term ${\text{d} \mathcal{L}} / {\text{d} \boldsymbol{\tilde{M}}}$ is used to guide the optimizer. The key difference lies in the microstructure descriptor. We propose to take an arbitrary microstructure descriptor which, if necessary, can be modified to make it differentiable as discussed in Section~\ref{sec:corrs}. In contrast, a fixed choice of the microstructure descriptor $\boldsymbol{G}^l$ is made in~\cite{li_transfer_2018}, where only slight modifications are required as $\boldsymbol{G}^l$ is already essentially differentiable\footnote{Max-pooling layers were changed to average-pooling layers to achieve better derivative properties. Note, however, that the Rectified Linear Units (ReLUs), despite being differentiable, may zero the gradient in certain cases.}. As can be seen, if $\boldsymbol{G}^l$ is used as a descriptor in the differentiable MCR concept (see Section~\ref{sec:theory}), deep CNNs can be viewed as a special case.

Having identified deep CNN synthesis~\cite{li_transfer_2018} as a particular form of differentiable MCR, we revisit the local minimum problem. 
As explained in Section~\ref{sec:formulation}, the presence or absence of local minima in differentiable MCR depends on the choice of the differentiable characterization function~$\boldsymbol{\tilde{f}_C}$.
Choosing~$\boldsymbol{\tilde{f}_C}$ as Gram matrices leads to the specific case of deep CNN synthesis. This unique Gram matrices based~$\boldsymbol{\tilde{f}_C}$ is shown to be capable of reconstructing a variety of microstructures with impressive results, even with multiple phases and in three dimensions~\cite{li_transfer_2018, bostanabad_reconstruction_2020}. 
Thus, besides the theoretical considerations in Section~\ref{sec:formulation} that the local minimum problem should be solvable for differentiable MCR, the success of deep CNN synthesis provides empirical evidence that, with a suitable descriptor, local minima can indeed be avoided.

\section{Numerical experiments}
\label{sec:exp}
The implementation details of the proposed formulation (Section~\ref{sec:formulation}) are given in Appendix~\ref{sec:implementationaldetails} which considers the spatial correlations discussed in Section~\ref{sec:corrs} as a microstructure descriptor. To assess the accuracy and the performance of the proposed framework, reconstruction results of various numerical experiments are presented. The analyses begin with a classical periodic two-phase checkerboard structure. For this experiment, the reconstruction results obtained with differentiable MCR and the Yeong-Torquato algorithm are compared in terms of accuracy and computational cost. As a second example, a wide variety of complex heterogeneous structures are reconstructed. The main focus herein is to demonstrate the obtainable reconstruction accuracy for heterogeneous media with scalable computational effort.

\subsection{Checkerboard structure}
The checkerboard has become a standard test case in the MCR literature~\cite{li_review_2014}. Its clear arrangement and simple structure make it straightforward to estimate the predicted accuracy, the noise, as well as the performance of the various particular formulations. 

\begin{figure}[H]
    \centering
    \input{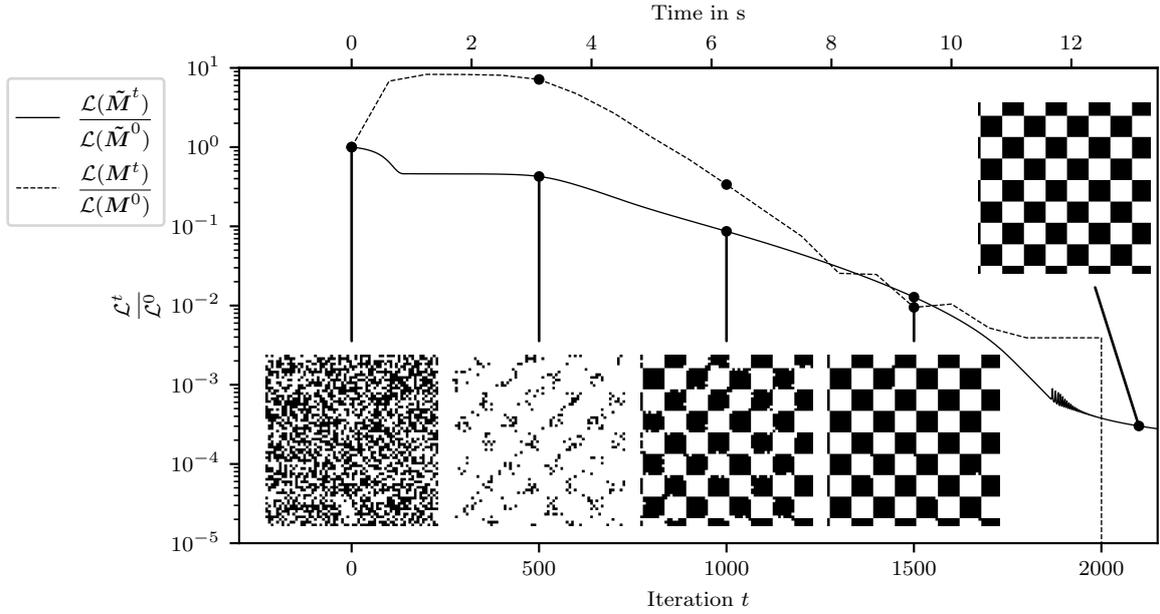}
    \caption{Differentiable MCR convergence plot for checkerboard structure of size $64 \times 64$ pixels. Besides the loss that is actually minimized, $\mathcal{L}(\boldsymbol{\tilde{M}}^t)$, 
    the loss of the rounded pixel values $\mathcal{L}(\boldsymbol{{M}}^t)$ and the associated intermediate results $\boldsymbol{{M}}^t$ are also shown.}
    \label{fig:convergence}
\end{figure}

For the domain size $64 \times 64$, Figure~\ref{fig:convergence} illustrates the convergence behavior of the proposed algorithm. The loss function $\mathcal{L}(\boldsymbol{\tilde{M}}^t)$ defined in~(\ref{L_func_dMCR}) is, except for a small unstable region, a non-increasing curve as expected. For comparability, in case of $\boldsymbol{\tilde{M}}^t$ rounded to obtain an integer-valued equivalent ${\boldsymbol{{M}} \in \{0, \, 1\}^{I \times J}}$, 
the associated loss $\mathcal{L}(\boldsymbol{M}^t)$ is plotted. This second loss curve increases significantly several times and, finally, after around $t = 2000$ iterations, $\mathcal{L}(\boldsymbol{M}^t)$ drops to zero\footnote{NB: When $\mathcal{L}(\boldsymbol{M}^t)$ drops to zero, the loss corresponding to the non-rounded structure, $\mathcal{L}(\boldsymbol{\tilde{M}}^t)$, is not zero yet, as the optimizer still needs to ``round'' the entries as well. This stage can be omitted.} as $\boldsymbol{M}$ becomes perfect.

\begin{figure}[H]
    \centering
    \input{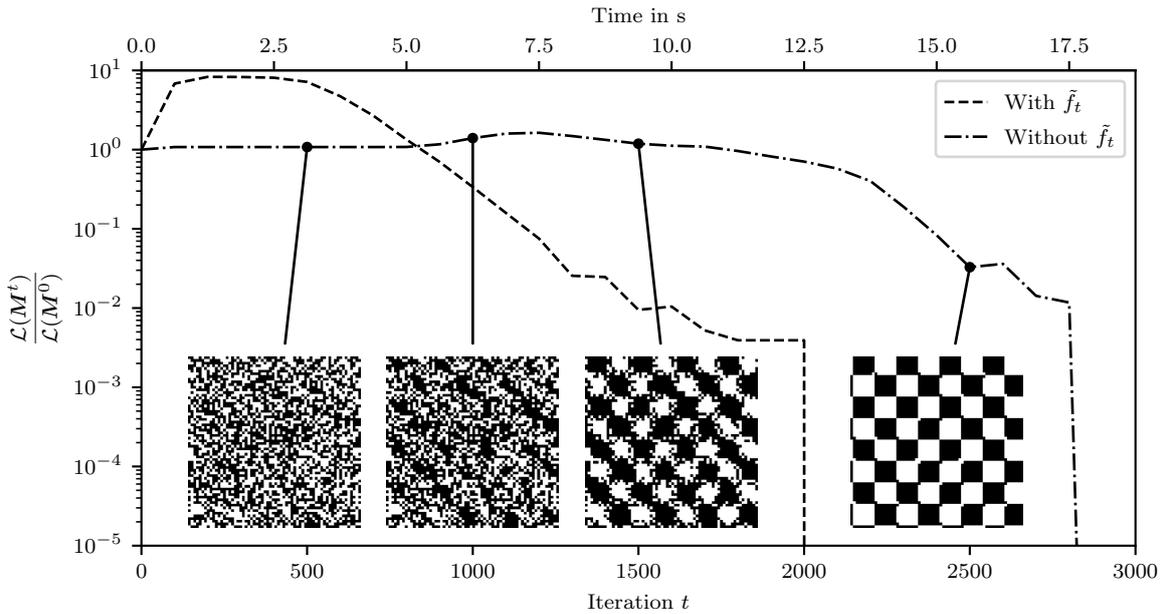}
    \caption{Illustration of the relevance of the asymmetric thresholding function~$\tilde{f}_t$ for a checkerboard structure of size $64 \times 64$ pixels. If it is omitted, the optimizer requires significantly more iterations in the first stage of convergence in order to find the optimal direction to start from the random initialization. This becomes especially clear through comparing the exemplary intermediate microstructures to those in Figure~\ref{fig:convergence}.}
    \label{fig:nothreshold}
\end{figure}

The exemplary intermediate results $\boldsymbol{M}^t$ depicted in Figure~\ref{fig:convergence} reveal the typical strategy of the optimizer. Starting with setting almost all pixels to phase 0, the volume fraction of phase 1 is gradually increased again, whereas only very few pixels are changed from phase 1 to 0 in the later stages of convergence. This asymmetry between phase 0 and 1 is due to the asymmetry of the soft threshold function~$\tilde{f}_t$. Recall that requiring $\tilde{f}_t'(x) > 0 \; \forall x \in (0, \, 1)$ led to $\tilde{f}_t(0.5) > 0$, what can be corrected as shown in Appendix~\ref{sec:correction}. To verify the asymmetry hypothesis, the soft threshold is replaced with $\tilde{f}_t(x) = x$, omitting the thresholding step\footnote{NB: The corrections derived in Appendix~\ref{sec:correction} still hold in this scenario.}, see Section~\ref{sec:s2stilde}. 
Figure~\ref{fig:nothreshold} presents the convergence behavior without and with thresholding function. Without $\tilde{f}_t$, as can be seen, the optimizer initially encounters difficulties in finding the optimal direction to start from the random initialization. Although the final result remains unaffected, in absence of the thresholding step the optimizer requires significantly more time for the initial stage of convergence, while the optimizer with thresholding function completes the initial phase within the first 500 iterations.
The same phenomenon is observed with a sine-wave threshold or other functions that are symmetric around 0.5. We conjecture that asymmetry in $\boldsymbol{\tilde{f}_C}$ is an essential cornerstone of efficient differentiable MCR, at least if based on the proposed differentiable correlations.

\begin{figure}[H]
    \centering
    \input{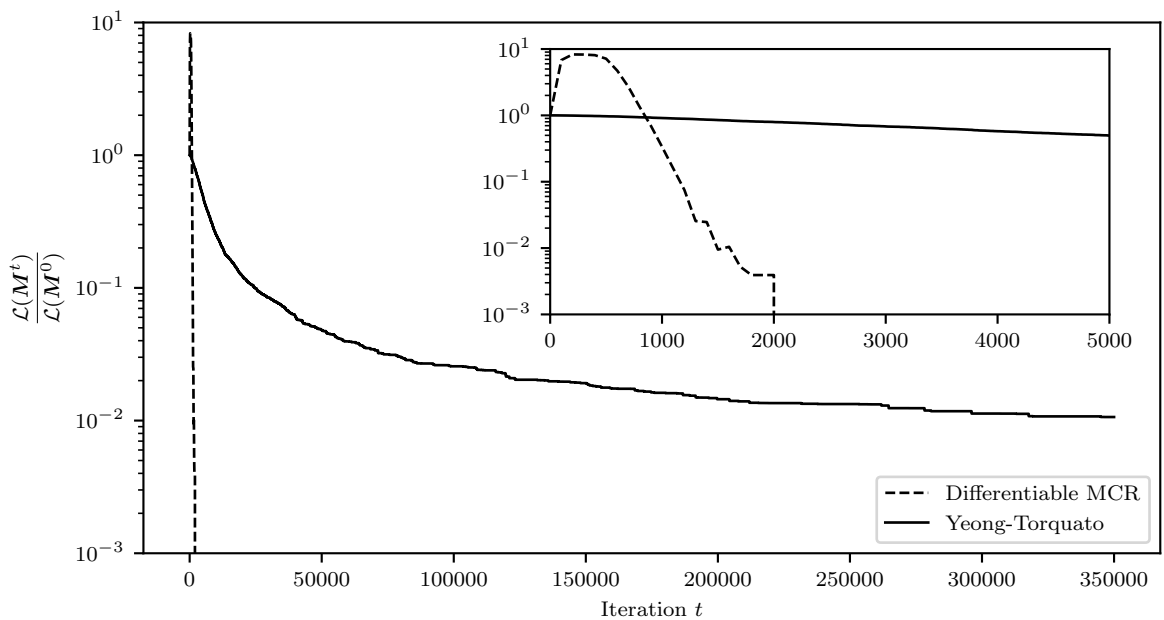}
    \caption{Convergence plot of the loss function of the Yeong-Torquato algorithm and differentiable MCR for reconstructing a checkerboard structure with~$64 \times 64$ pixels. Both results stem from the author's implementations. A zoom to the first few iterations is provided in the upper right corner.}
    \label{fig:convvsyt}
\end{figure}

Figure~\ref{fig:convvsyt} reports a convergence comparison between differentiable MCR and the author's implementation of the Yeong-Torquato algorithm\footnote{The author's implementation precisely follows Algorithm~\ref{alg:yt} with a cool-down factor of~$\alpha=0.9$ and proposes pixel swaps between the two phases completely randomly. Note that numerous significantly more efficient variants exist.} for a checkerboard structure with $64 \times 64$ pixels. The performance of the Yeong-Torquato algorithm drops drastically in the later stages of convergence\footnote{To obtain the error~(\ref{L_func_YT}) close to zero, the same Yeong-Torquato implementation was run for an entire week ($10^7$ iterations, not shown here). With only two ``wrong'' pixels remaining in the end, the error could be reduced almost to zero, but not completely. Because only two pixels were left to be swapped, the algorithm was clearly not trapped in a local minimum. Therefore, this can not be explained by an overly fast cool-down, but only by the algorithm itself.}. As discussed previously, this behavior is inherent to its stochasticity, because as the loss decreases, randomly proposed pixel swaps become increasingly unlikely to ameliorate the result. On the contrary, the gradient-based optimizer of differentiable MCR does not stagnate at low errors. It continues to decrease the error quickly, yielding reconstruction result that show \emph{exact} statistical equivalence with the original structure. Importantly, it can be seen that differentiable MCR requires significantly fewer iterations than the Yeong-Torquato algorithm\footnote{It should be added that the necessity to compute a gradient makes each iteration in differentiable MCR more computationally expensive than an iteration of the Yeong-Torquato algorithm, but this aspect does not go near the differences shown in Figure~\ref{fig:convvsyt}.}. 

\begin{figure}[H]
    \centering
    \input{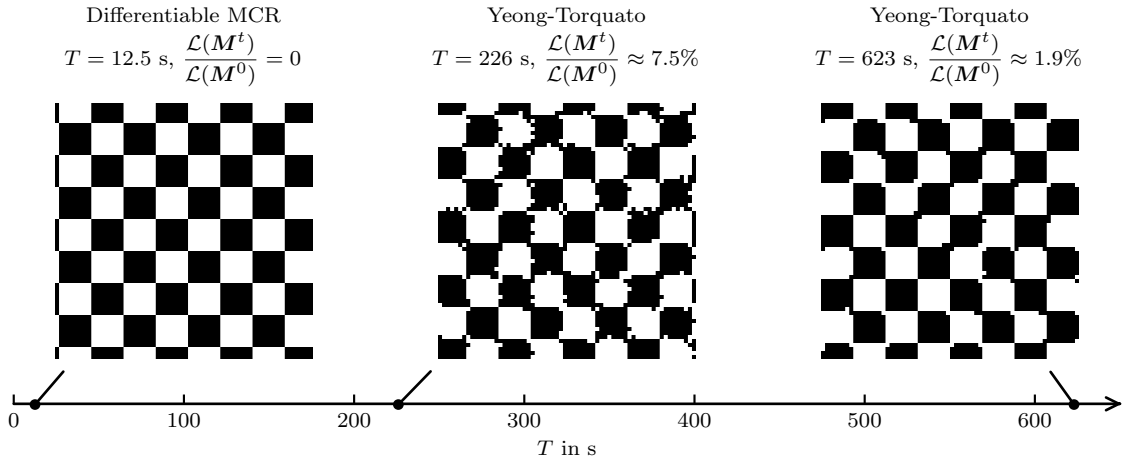}
    \caption{Comparison between performance of Yeong-Torquato algorithm and differentiable MCR in reconstructing a checkerboard structure of size $64 \times 64$ pixels. The results for the former stem from a well-optimized implementation given in~\cite{li_review_2014, li_comparison_2012}. Reproduced with permission from the publisher.}
    \label{fig:diffmcrvsyt}
\end{figure}

In general, a direct efficiency comparison between differentiable MCR and the Yeong-Torquato algorithm is difficult. It does not only depend on the implementation platform, the hyper-parameters and which of the many variants is chosen, but also strongly on the desired accuracy. 
In order to reasonably compare the wall-clock time of the proposed method with the Yeong-Torquato algorithm, results from a well-optimized version are taken from the literature~\cite{li_review_2014, li_comparison_2012}. For the same checkerboard example, Figure~\ref{fig:diffmcrvsyt} reports wall-clock times for the Yeong-Torquato algorithm at different levels of accuracy. After less than $4$ minutes, the checkerboard pattern can be recognized, but with significant noise. After more than $10$ minutes, the noise is reduced but still remains. Differentiable MCR requires $12.5$ s for a perfect reconstruction. 
According to Figures~\ref{fig:convvsyt} and~\ref{fig:diffmcrvsyt}, the performance gain is one order of magnitude for an error of around $10 \%$, whereas for lower or almost-zero errors, differentiable MCR is many orders of magnitude faster. 
We emphasize that the reason behind this acceleration is the differentiable optimization scheme the proposed approach is based on.

\begin{figure}[H]
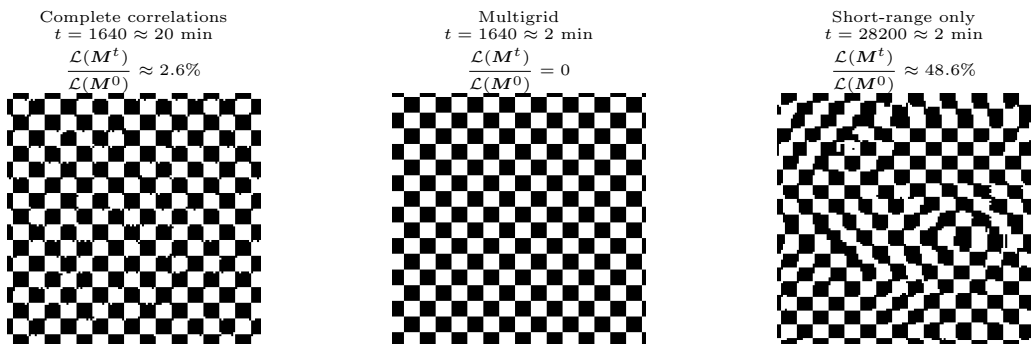

    \centering
    \input{recms_mg_full.pgf}
    \hspace{1cm}
    \input{recms_mg_normal.pgf}
    \hspace{1cm}
    \input{recms_mg_nomg.pgf}
    \caption{Comparison between performance of the multigrid methods for checkerboard structure of size $128 \times 128$ pixels. Left: descriptor contains correlations corresponding to all vectors, also long-range vectors; right: descriptor is limited to correlations corresponding to short-range vectors only; center: proposed multigrid scheme.}
    \label{fig:reconstructionsmg}
\end{figure}

\begin{figure}[H]
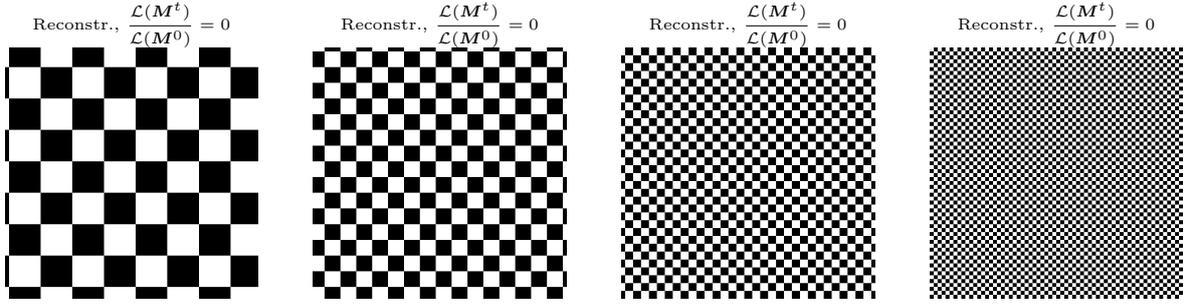

    \centering
    \input{recms_c_64_0.pgf}
    \hfill
    \input{recms_c_128_00.pgf}
    \hfill
    \input{recms_c_256_00.pgf}
    \hfill
    \input{recms_c_512_00.pgf}
    \caption{Reconstructing checkerboard structures of sizes $I, J \in \{64, \, 128, \, 256, \, 512 \}$ pixels with differentiable MCR using proposed multigrid scheme.}
    \label{fig:checerboardreconstructions}
\end{figure}

With the chosen maximum correlation length of $P=Q=31$, see Appendix~\ref{sec:implementationaldetails}, for the case of a $64\times 64$ pixel structure, the multigrid strategy discussed in Section~\ref{sec:mg} is not needed. As the microstructure resolution increases, however, computing all long-range correlations quickly becomes computationally expensive. Figure~\ref{fig:reconstructionsmg} reports the performance gains of the multigrid method for the structures with the domain size $128 \times 128$. This proposed method requires around $2$ minutes and less than~$2000$ iterations to obtain the exact reconstruction. In case of limiting the descriptor to short-range\footnote{For illustration purposes, we set $P=Q=8$ \emph{only} for this example and \emph{only} for the short-range version. The same behavior can be observed with $P=Q=32$, but less pronounced.} correlations, more iterations are required in the same time to correctly match the global correlations. The long-range version requires high computational effort. The proposed technique yields a compromise to include both highly-accurate short-range and less accurate long-range information. Figure~\ref{fig:checerboardreconstructions} shows well-reconstructed checkerboard structures at different resolutions.

\begin{figure}[H]
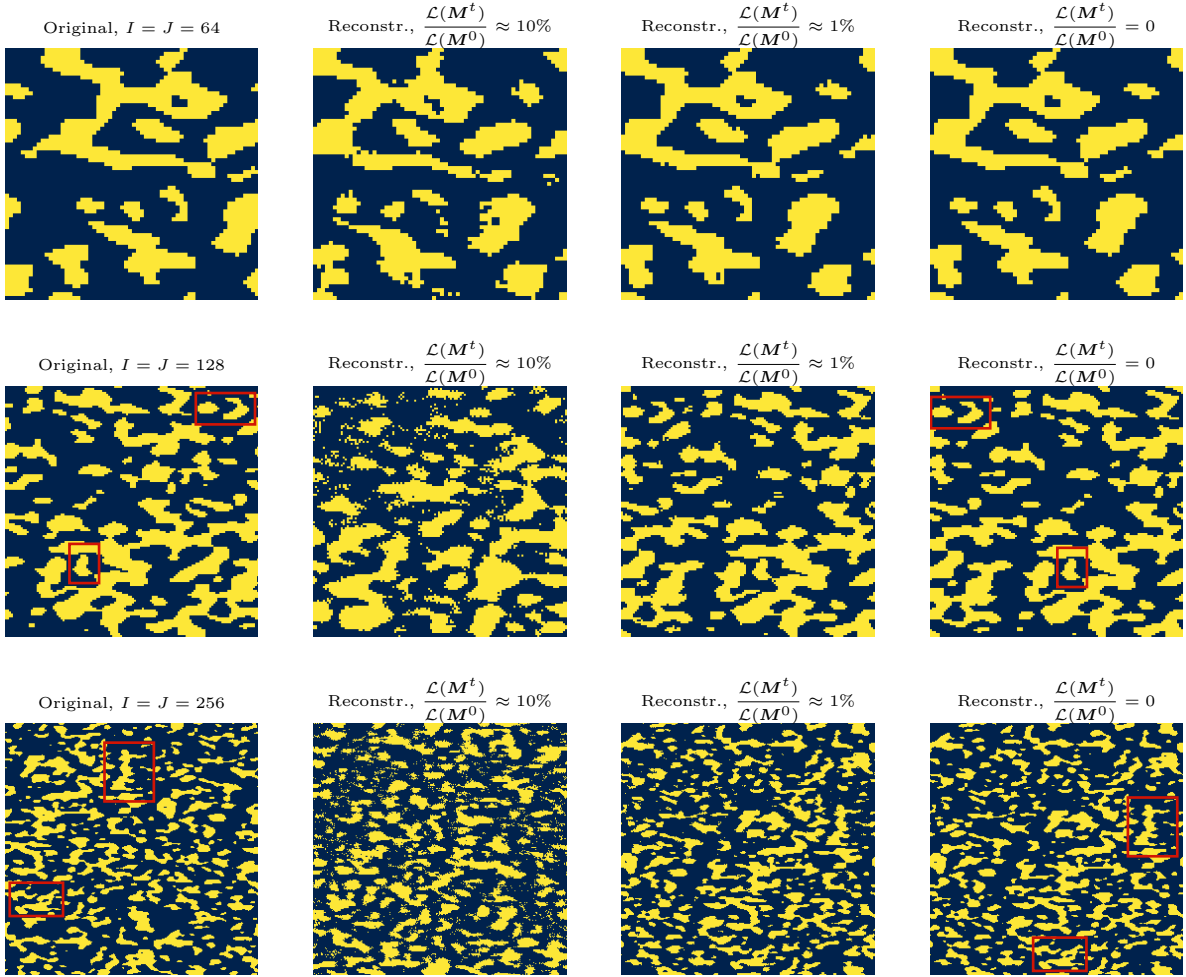

    \centering
    \input{ogms_64.pgf}
    \hfill
    \input{recms_64_10.pgf}
    \hfill
    \input{recms_64_01.pgf}
    \hfill
    \input{recms_64_00.pgf}
    \vfill
    \input{ogms_128.pgf}
    \hfill
    \input{recms_128_10.pgf}
    \hfill
    \input{recms_128_01.pgf}
    \hfill
    \input{recms_128_00.pgf}
    \vfill
    \input{ogms_256.pgf}
    \hfill
    \input{recms_256_10.pgf}
    \hfill
    \input{recms_256_01.pgf}
    \hfill
    \input{recms_256_00.pgf}
    \caption{Differentiable MCR results for reconstructing synthetically generated heterogeneous structures~\cite{brough_materials_2017}. The corresponding wall-clock times are presented in Figure~\ref{fig:scaling}. In the higher-resolution cases, a characteristic feature has been highlighted through a red box to show that the reconstructions are exact translations of the original microstructure.}
    \label{fig:reconstructions}
\end{figure}
\subsection{Random media reconstruction}
\begin{figure}[H]
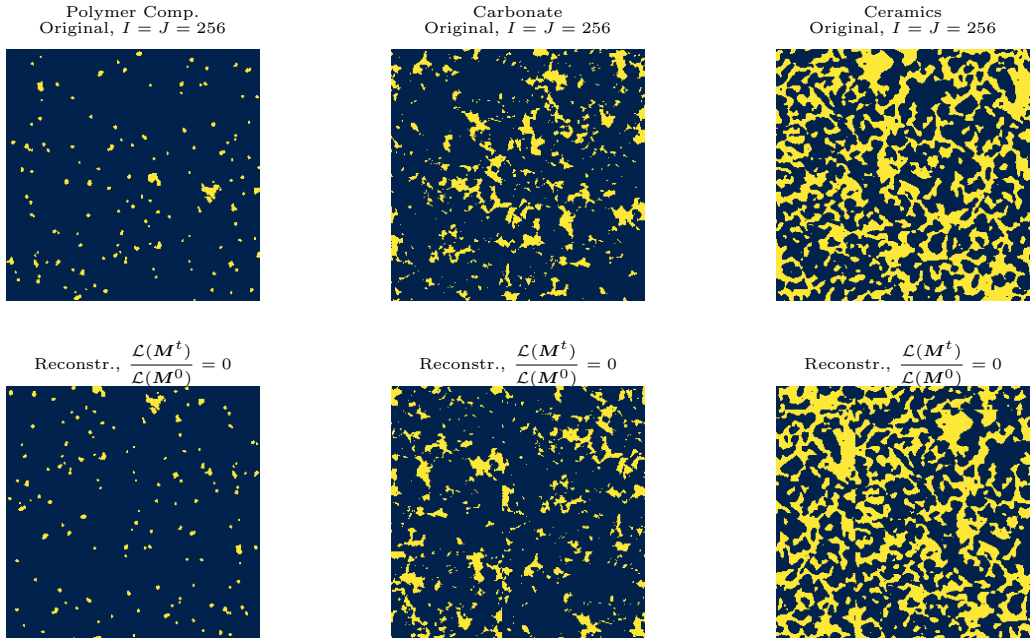

    \centering
    \input{ogms_PMMASiO2.pgf}
    \hspace{1cm}
    \input{ogms_carbonate.pgf}
    \hspace{1cm}
    \input{ogms_ceramics.pgf}
    \vfill
    \input{recms_PMMASiO2.pgf}
    \hspace{1cm}
    \input{recms_carbonate.pgf}
    \hspace{1cm}
    \input{recms_ceramics.pgf}
    \caption{Differentiable MCR results for reconstructing different microstructures presented by Li~et~al.~\cite{li_transfer_2018}, re-used with kind support from the authors and permission from the publisher through the \emph{Creative Commons} license~\cite{cc_licence}.}
    \label{fig:lireconstructions}
\end{figure}

\begin{figure}[H]
    \centering
    \input{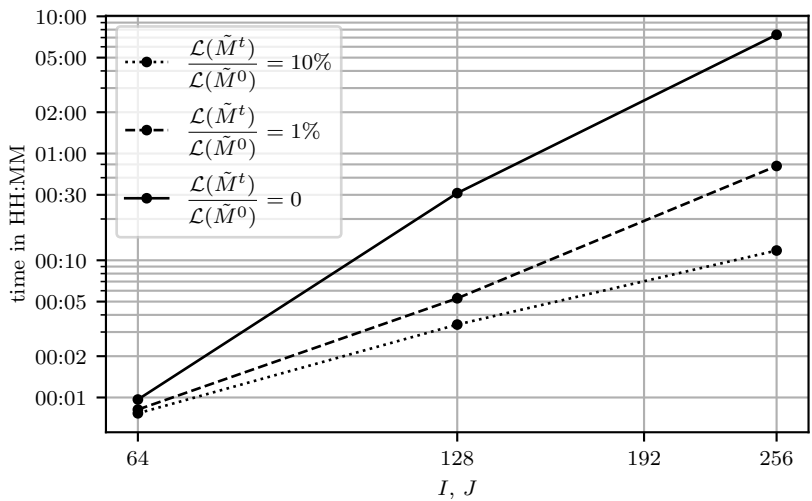}
    \caption{Scaling behavior of differentiable MCR for synthetically generated heterogeneous structures~\cite{brough_materials_2017} shown in Figure~\ref{fig:reconstructions}. 
    }
    \label{fig:scaling}
\end{figure}

In this example, the reconstruction performance of differential MCR is demonstrated for realistic microstructures. To begin with, synthetic heterogeneous structures generated using \emph{pyMKS}~\cite{brough_materials_2017} for the chosen domain sizes of $64 \times 64$, $128 \times 128$ and $256 \times 256$ pixels. The original and reconstructed heterogeneous structures at different stages of convergence are shown in Figure~\ref{fig:reconstructions}. In addition to this, real microstructures from~\cite{li_transfer_2018} are reconstructed and shown in Figure~\ref{fig:lireconstructions}. Even with significantly more complex features in these microstructures, the original statistical characteristics are captured exactly in the reconstructed image, as the final result is nothing but a translated version of the original image. This holds not only for the synthetically generated periodic microstructures in Figure~\ref{fig:reconstructions}, but even for the non-periodic real data in Figure~\ref{fig:lireconstructions}. The computational cost associated with each image in Figure~\ref{fig:reconstructions} is reported in Figure~\ref{fig:scaling}. These findings are in line with that of the previous example, except that reconstructing the synthetic microstructure takes more time than the checkerboard structure. The computational effort of the algorithm thus depends on both the choice of $\boldsymbol{\tilde{f}_C}$ and its desired value $\boldsymbol{D}^{\text{des}}$. Furthermore, note the relatively straight lines in the double-logarithmic plot in Figure~\ref{fig:scaling}. This indicates a relation between the time~$T$ and the linear size~$I$ of the microstructure of $\ln T \propto \ln I$. In this case, $T = I^c$, where $c$ depends on the desired resolution. The general validity of this relation must, however, be confirmed through further investigations. 


\section{Conclusions and Outlook}
\label{sec:summary}
In this contribution, differentiable microstructure characterization and reconstruction is proposed. Like the classical Yeong-Torquato algorithm, it shares the central idea that microstructure reconstruction is optimization. 
The novelty of this work is motivated by recognizing that, despite ever better heuristics, the Yeong-Torquato algorithm is essentially bound to moving randomly through the inconceivably large space of possible microstructures. 
This randomness reduces the performance, especially at the later stages of convergence.
Thus, the concept of differentiable MCR is formulated in depth along with the requirements it brings for the descriptor and the microstructure. The difficulties thereby lie in formulating differentiable descriptors and avoiding local minima. We argue that the latter can be addressed by the former and provide empirical evidence through identifying a successful existing approach as a special case of differentiable MCR. Furthermore, spatial two- and three-point correlations are recast as a suitable differentiable descriptor. 
This specific choice of descriptor defines one of many versions of differentiable MCR.
In principle one can choose \emph{any} suitable differentiable descriptor or even combine multiple descriptors. 

The presented correlation-based version of differentiable MCR naturally runs on a GPU. Its scalability is ensured by a proposed multigrid scheme.
Although the code is far from fully optimized, \emph{perfect} reconstruction results are achieved in relatively short time through extremely fast convergence. In contrast, existing algorithms can equally well reduce the descriptor error norm by one or two orders of magnitude, but take incomparably more time to reduce it further. 
This difference supports the central hypothesis of this work: Microstructure reconstruction is optimization and efficient optimization requires the gradient.
Overall, differentiable MCR, due to its flexibility, controllability and no requirement of training data, seems to be a unified mathematically sound framework for reconstructing random heterogeneous media. It may serve as a useful tool for establishing PSP-linkages and realizing the inverse material design concept. To achieve real-world applicability, however, several additional aspects have to be investigated. 
\begin{enumerate}
    \item The applicability of differentiable MCR to more complex situations should be validated or established. This includes microstructures with more than two phases, with more than one field (e.g. phase and orientation) and in three dimensions.
    \item Additional descriptors should be developed. In principle, any descriptor should be generalizable to a differentiable version, but doing so is not always straightforward and involves several engineering decisions. Furthermore, different descriptors may be combined in a single loss function.
    \item The implementation of differentiable MCR should be optimized for performance. For example, this includes accounting for the sparsity of the convolution masks in the proposed correlation descriptor. More generally, it also involves reformulating existing descriptors such that they are efficiently computable, like it has been done for the Yeong-Torquato algorithm. Finally, the optimizer itself may also be subject to further research.
\end{enumerate}
With the presented basis and the possible extensions, we wish to contribute to speeding up materials discovery, a key enabler for engineering progress in many fields.

\section*{Acknowledgements}
This research is funded by the European Regional Development Fund (ERDF) and co-financed by tax funds based on the budget approved by the members of the Saxon State Parliament.
\begin{figure}[H]
    \centering
    \includegraphics[width = 0.5\linewidth]{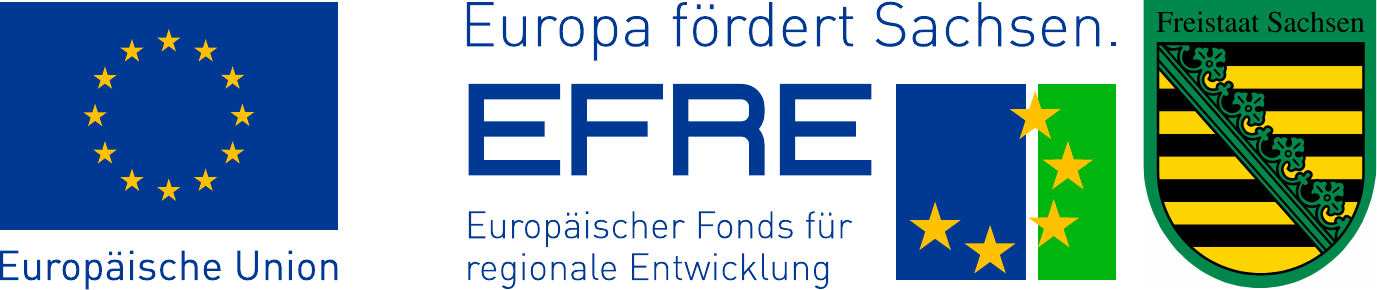}
\end{figure}
The authors are grateful to the Centre for Information Services and High Performance Computing [Zentrum für Informationsdienste und Hochleistungsrechnen (ZIH)] TU Dresden for providing its facilities for high throughput calculations.



\section*{Competing interests}
The authors declare no competing interests.

\section*{Author contributions}
P.S., M.A. and A.R. developed the concepts under coordination of M.K. and P.S. conducted the computations.
All authors contributed to the discussions, analyzed the simulation results and wrote the manuscript.


\appendix

\section{Correlation correction for increasing soft threshold}
\label{sec:correction}
In this Appendix, due to the introduced soft threshold function in Section~\ref{sec:s2stilde}, we investigate the resulting correction for both two-point $\boldsymbol{S}_2^{1 \to 1}(\vec{r})$ and three-point $\boldsymbol{S}_3^{1 \to 1}(\vec{r}_a, \, \vec{r}_b)$ correlations. 

For the case of $\boldsymbol{M} \in \{0, \, 1\}^{I \times J}$, the discrepancy between $\boldsymbol{\bar{S}}_2^{1 \to 1}(\vec{r})$ and $\boldsymbol{S}_2^{1 \to 1}(\vec{r})$ stems from the contributions of the reduction where only one end of $\vec{r}$ is in phase one. These contributions are weighted with $\tilde{f}_t(\nicefrac{1}{2})$. They are essentially cross-correlations $\boldsymbol{S}_2^{0 \to 1}(\vec{r})$ and $\boldsymbol{S}_2^{1 \to 0}(\vec{r})$
\begin{equation}
    \boldsymbol{\bar{S}}_2^{1 \to 1}(\vec{r}) = \boldsymbol{S}_2^{1 \to 1}(\vec{r}) +  \tilde{f}_t^2(\nicefrac{1}{2}) \boldsymbol{S}_2^{0 \to 1}(\vec{r}) + \tilde{f}_t^2(\nicefrac{1}{2}) \boldsymbol{S}_2^{1 \to 0}(\vec{r}).
\end{equation}
From the definition of the two-point correlations it follows that
\begin{equation}
\boldsymbol{S}_2^{0 \to 1}(\vec{r}) = \boldsymbol{S}_2^{1 \to 0}(\vec{r}) = \boldsymbol{S}_2^{1 \to 1}(\vec{0}) - \boldsymbol{S}_2^{1 \to 1}(\vec{r})    
\end{equation}
where $\boldsymbol{S}_2^{1 \to 1}(\vec{0})$ is the volume fraction of phase 1. This leads to
\begin{equation}
\begin{aligned}
\boldsymbol{\bar{S}}_2^{1 \to 1}(\vec{r})
& =  \left\{
\begin{array}{ll}
\boldsymbol{S}_2^{1 \to 1}(\vec{r}), & \vec{r} = \vec{0} \\
(1-2\tilde{f}_t^2(\nicefrac{1}{2})) \cdot \boldsymbol{S}_2^{1 \to 1}(\vec{r}) + 2 \tilde{f}_t^2(\nicefrac{1}{2}) \cdot \boldsymbol{S}_2^{1 \to 1}(\vec{0}) , & \vec{r} \neq \vec{0} \\
\end{array}
\right. .\\
\end{aligned}
\end{equation}
Inverting the mapping allows to express $\boldsymbol{S}_2^{1 \to 1}(\vec{r})$ only in terms of $\boldsymbol{\bar{S}}_2^{1 \to 1}(\vec{r})$, as
\begin{equation}
\boldsymbol{S}_2^{1 \to 1}(\vec{r}) =  \left\{
\begin{array}{ll}
\boldsymbol{\bar{S}}_2^{1 \to 1}(\vec{r}), & \vec{r} = \vec{0} \\
\dfrac{\boldsymbol{\bar{S}}_2^{1 \to 1}(\vec{r}) - 2 \tilde{f}_t^2(\nicefrac{1}{2}) \cdot \boldsymbol{\bar{S}}_2^{1 \to 1}(\vec{0}) }{1 - 2 \tilde{f}_t^2(\nicefrac{1}{2})}, & \vec{r} \neq\vec{0} \\
\end{array}
\right. .\\
\label{eqn:s2rfroms2rbar}
\end{equation}
The mapping~(\ref{eqn:s2rfroms2rbar}) is derived to yield the original two-point correlations for $\boldsymbol{M} \in \{0, \, 1\}^{I \times J}$. For the case of $\boldsymbol{\tilde{M}} \in \mathbb{R}^{I \times J}$, the same expression ~(\ref{eqn:s2rfroms2rbar}) can be used to obtain a differentiable microstructure descriptor $\boldsymbol{\tilde{S}}_2^{1 \to 1}(\vec{r})$ that satisfies the requirements discussed in Section~\ref{sec:formulation}
\begin{equation}
\boldsymbol{\tilde{S}}_2^{1 \to 1}(\vec{r}) =  \left\{
\begin{array}{ll}
\boldsymbol{\bar{S}}_2^{1 \to 1}(\vec{r}), & \vec{r} = \vec{0} \\
\dfrac{\boldsymbol{\bar{S}}_2^{1 \to 1}(\vec{r}) - 2 \tilde{f}_t^2(\nicefrac{1}{2}) \cdot \boldsymbol{\bar{S}}_2^{1 \to 1}(\vec{0}) }{1 - 2 \tilde{f}_t^2(\nicefrac{1}{2})}, & \vec{r} \neq\vec{0} \\
\end{array}
\right. \\
\end{equation}

Following a similar line of argumentation, $\boldsymbol{S}_3^{1 \to 1}(\vec{r}_a, \, \vec{r}_b)$ is obtained from $\boldsymbol{\bar{S}}_3^{1 \to 1}(\vec{r}_a, \, \vec{r}_b)$. In this case, the discrepancy between $\boldsymbol{\bar{S}}_3^{1 \to 1}(\vec{r}_a, \, \vec{r}_b)$ and $\boldsymbol{S}_3^{1 \to 1}(\vec{r}_a, \, \vec{r}_b)$ stems from the following contributions, where
\begin{itemize}
    \item both vectors $\vec{r}_a$ and $\vec{r}_b$ start in phase 1 and end in phase 0, weighted with $\tilde{f}_t^2(\nicefrac{1}{2})$, with a probability of $\boldsymbol{S}_2^{1 \to 1}(\vec{0}) - \boldsymbol{S}_2^{1 \to 1}(\vec{r}_a) - \boldsymbol{S}_2^{1 \to 1}(\vec{r}_b) + \boldsymbol{S}_3^{1 \to 1}(\vec{r}_a, \, \vec{r}_b)$,
    \item both vectors $\vec{r}_a$ and $\vec{r}_b$ start in phase 0 and end in phase 1, weighted with $\tilde{f}_t^2(\nicefrac{1}{2})$, with a probability of $\boldsymbol{S}_2^{1 \to 1}(\vec{r}_a - \vec{r}_b) - \boldsymbol{S}_3^{1 \to 1}(\vec{r}_a, \, \vec{r}_b)$, and
    \item both vectors $\vec{r}_a$ and $\vec{r}_b$ start in phase 1, but only one ends in phase 1, weighted with $\tilde{f}_t(\nicefrac{1}{2})$, with a probability of $\boldsymbol{S}_2^{1 \to 1}(\vec{r}_a) + \boldsymbol{S}_2^{1 \to 1}(\vec{r}_b) - 2 \boldsymbol{S}_3^{1 \to 1}(\vec{r}_a, \, \vec{r}_b)$.
\end{itemize}
Exploiting that $\boldsymbol{S}_2^{1 \to 1}(\vec{r})$ can be obtained from $\boldsymbol{\bar{S}}_2^{1 \to 1}(\vec{r})$ and that $\boldsymbol{S}_2^{1 \to 1}(\vec{r})$ is included in $\boldsymbol{S}_3^{1 \to 1}(\vec{r}_a, \, \vec{r}_b)$ for $\vec{r}_a = \vec{r}_b$ allows to define a mapping 
\begin{equation}
    \boldsymbol{S}_3^{1 \to 1}(\vec{r}_a, \, \vec{r}_b) = \left\{ 
    \begin{array}{ll}
    \boldsymbol{{S}}_2^{1 \to 1}(\vec{r}_a), & \vec{r}_a = \vec{r}_b \\
    \boldsymbol{{S}}_2^{1 \to 1}(\vec{r}_a), & \vec{r}_a \neq \vec{0}, \, \vec{r}_b = \vec{0} \\
    \boldsymbol{{S}}_2^{1 \to 1}(\vec{r}_b), & \vec{r}_a = \vec{0}, \, \vec{r}_b \neq \vec{0} \\
    \bigg[ \boldsymbol{\bar{S}}_3^{1 \to 1}(\vec{r}_a, \, \vec{r}_b) - \tilde{f}_t(\nicefrac{1}{2}) \left( \boldsymbol{S}_2^{1 \to 1}(\vec{r}_a) + \boldsymbol{S}_2^{1 \to 1}(\vec{r}_b)  \right) - ... & \\
    \qquad \tilde{f}_t^2(\nicefrac{1}{2}) \big( \boldsymbol{S}_2^{1 \to 1}(\vec{0}) - \boldsymbol{S}_2^{1 \to 1}(\vec{r}_a) - \boldsymbol{S}_2^{1 \to 1}(\vec{r}_b) + ... & \\
    \qquad \boldsymbol{S}_2^{1 \to 1}(\vec{r}_a - \vec{r}_b) \big) \bigg] \cdot \dfrac{1}{1 - 2\tilde{f}_t(\nicefrac{1}{2})} & \vec{0} \neq \vec{r}_a \neq \vec{r}_b \neq \vec{0} \\
\end{array}
    \right.
\label{eqn:s3rfroms3rbar}
\end{equation}
which depends only on known quantities.

\section{Implementational Details}
\label{sec:implementationaldetails}
We implemented the application of differentiable MCR to spatial correlations presented in Section~\ref{sec:corrs}. We have choosen $\boldsymbol{\bar{S}}^{1\to 1}$ as a microstructure descriptor, not $\boldsymbol{{S}}^{1\to 1}$, omitting (~\ref{eqn:s2rfroms2rbar}) and~(\ref{eqn:s3rfroms3rbar}) for simplicity, as they have been shown to contain the same information. The multigrid approach presented in Section~\ref{sec:mg} is implemented, where the maximum correlation length is set to $P=Q=31$, resulting in no multigrid for $I=J=64$, two levels for $I=J=128$ and three for $I=J=256$. The point symmetry of the ensemble with respect to $\vec{r}$ is exploited to reduce the size of $\boldsymbol{m}$. Furthermore, masks corresponding to $\vec{r}$ with negative components are shifted into the positive region, allowing for convolution kernels with $32 \times 32$ entries. The consequent shift of the affected ensembles is corrected in an additional step by shifting them back again.

TensorFlow~\cite{abadi_tensorflow_2016} is chosen as the overarching programming environment for its automatic gradient computation, for compiling from Python to efficient GPU code and for its implementation of the chosen optimizer. 
We use the ADAM optimizer~\cite{kingma_adam_2017} with its standard TensorFlow hyperparameters $\alpha = 0.001$, $\beta_1 = 0.9$ and $\beta_2 = 0.999$. Furthermore, the code runs on the \emph{NVidia V100} GPU but only one of the many workers is used. Note that the utilized code is far from fully optimized, as this is not the focus of the present work.
\printbibliography

\end{document}